\documentstyle[twocolumn,aps]{revtex}
\begin{document}

\title{
Van Hove Exciton-Cageons and High-T$_c$ Superconductivity: XB: \\
Polaronic Coupling in the Doped Material}

\author{R.S. Markiewicz}

\address{Physics Department and Barnett Institute,
Northeastern U.,
Boston MA 02115}

\maketitle

\begin{abstract}
A purely ionic interpretation of the tilting mode instabilities in La$_{2-x}$A$
_x$CuO$_4$ (A=Sr,Ba) is shown to be not self-consistent:
the dominant factor influencing the doping dependence of the interlayer
mismatch
is the large change in the Cu-O bond length.  But such a large dependence of
bond length on valence leads to a strong electron-phonon coupling,
contradicting the assumption that covalent effects were negligible.  This
coupling is closely related to the vHs-JT effect.
This new insight clarifies the role of
the tilt-mode instabilities.  The main JT coupling is {\it not} to these modes,
but to the in-plane O-O bond stretching modes which split the vHs degeneracy.
However, as these modes soften, they couple to the lower-lying tilt modes, so
that the ultimate instability has a finite tilt component.
\par
This in turn simplifies the description of the JT instabilities.  The tilt
modes, while strongly coupled to the electrons, are quadratic in the tilt
angle.
On the other hand, the bond stretch modes have a large, linear coupling to
electrons, with clear polaronic effects.  This is connected with the fact that,
near a vHs, valence fluctuations are slow and relatively long lived, coupling
to
local phonon modes.  A striking result of this is that there will be a large
polaronic band narrowing near the vHs, whether or not the vHs is near the Fermi
level.  This vHs-localized band narrowing provides a natural explanation for
the
common occurence of extended vHs's.

\end{abstract}


\narrowtext

\section{Introduction}

Recent photoemission studies have demonstrated that a van Hove singularity
(vHs)
falls very close to the Fermi level in optimally doped Bi$_2$Sr$_2$CaCu$_2$O$_
8$ (Bi-2212)\cite{PE1} and YBa$_2$Cu$_3$O$_7$ (YBCO)\cite{PE2}.  This lends
strong support to the many proposals that the vHs plays an important role in
high-T$_c$ superconductivity\cite{vHs} in these materials.  The vHs's also
couple to many of the phonon modes involved in the various structural phase
transitions found in La$_{2-x}$Sr$_x$CuO$_4$ (LSCO) and related
compounds\cite{RM8A,RM8B,RM8C,RM8}, which leads to a competition between
superconductivity and structural transitions in the cuprates.
\par
However, these structural phase transitions are common to most perovskites,
many
of which are insulating, so the notion that electron-phonon coupling plays an
important role has been questionned.  The conventional view of these
transitions
has been that they are purely ionic, driven by the Madelung energy or
interlayer
strain mismatch, and have little or nothing to do with the electronic system.
\par
This conventional picture has been modified by recent analyses.  The
perovskites
have two classes of distortions, ferroelectric (FE), which generate a dipole
moment in the unit cell, and antiferrodistortive (AFD), involving tilting of
local octahedra.  Cohen\cite{Coh} has shown that there can be large {\it
covalent} contributions to the FE instabilities, {\it even when the materials
are insulating}, if the Fermi level lies in a hybridization gap between two
atomic species. (A similar result is discussed in Appendix A4 of Ref.
\cite{RM8B}).  This has stimulated a great deal of research showing the
important role of correlation effects\cite{Post} in FE instabilities.  Of
particular interest is a recent two-band Hubbard model of the FE instability,
introduced by Egami, et al.\cite{Eg}, which involves a one-dimensional (1D)
version of the three-band Hubbard model regularly applied to the cuprates.
\par
The structural phase transitions in the cuprates are generally AFD, involving
tilting of the CuO$_6$ octahedra.  Such
AFD instabilities are common in other perovskites, and indeed often
compete with FE instabilities\cite{ZhoVa}.  However, these AFD instabilities
are
generally found to be ionic rather than covalent -- the tendency towards
instability scales linearly with a tolerance factor, $t$, which measures the
degree of ionic size mismatch\cite{ZhoVa} (see, e.g., Eq. 1, below).  Hence,
the
present paper provides a reassessment of the role of electron-phonon coupling
in the structural instabilities of the cuprates.
\par
The paper is organized around three calculations.  First, a mixed
ionic-covalent model of the structural transitions is recalled.  It is shown
that the same model which explained the $T$-dependence of the low temperature
orthorhombic (LTO) phase in LSCO\cite{RM8B} can also explain its doping
dependence (Section II and Appendix A).  [Strictly speaking, the dominant
instability is to a phase of short-range order -- identified with the pseudogap
phase -- but the transition temperatures for this phase and the LTO phase are
proportional to each other.]  While the model {\it must} contain a significant
ionic contribution, there can also be a significant covalent contribution.
Secondly, the purely ionic model for the cuprates is analyzed in more detail,
and a number of discrepancies are pointed out.  Most importantly, it is found
that the dominant factor comtrolling the doping dependence of the LTO
transition
is the large compressibility of the CuO$_2$ planes: in ionic terms, there is a
large decrease in the Cu-O distance when $O^{2-}\rightarrow O^-$.  This
analysis
is presented in Section III and Appendix B.
\par
This large change of interionic spacing with valence is a signature of large
electron-phonon coupling, as found in rare earth compounds\cite{Hews}.  Hence,
a purely ionic model with no electron-phonon coupling is not self-consistent.
The contraction of the O-radius couples to a number of phonon modes, but
predominantly to in-plane modes, and not to tilting modes.
This leads to a reanalysis of earlier results.  It is now
suggested that the major phonon coupling is indeed with planar modes, which
couple quadratically to the much lower frequency octahedral tilt modes.  This
weak coupling ensures that when the highly coupled bond stretching modes start
to soften, the tilt modes go unstable first.
\par
This has important consequences for the vHs-JT effect.  The electron-tilt
phonon
coupling is quadratic in the tilt angle, and hence in the phonon creation
operator -- a strong violation of Migdal's theorem.  In contrast, the the O-O
bond stretching mode has a {\it linear JT coupling} to the vHs, via the length
dependence of the hopping parameter, $t_{CuO}$.  The soft mode behavior of, in
particular, the O-O stretch modes is discussed in Section IV and Appendix C.
Moreover, there is a clear {\it polaronic} contribution: as the hole hops
around, it carries this lattice distortion with it.  These polaronic effects
are
discussed in Section V and Appendix D.  By estimating the magnitude of the
polaronic corrections to the bandstructure, it will be possible to make
connection with various polaronic or bipolaronic theories of superconductivity.
\par
One striking result is that polaronic effects can explain the existence of
extended vHs's, {\it even when the vHs is away from the Fermi level.}
Band narrowing effects are localized near a vHs due to the fact that, near a
vHs, the material has an ioniccovalent crossover.  Electronic motion is slow
compared to phonon motion, so valence fluctuations are well defined (Section
VI). Section VII summarizes some conclusions, while Appendix E applies some of
these results to spin-orbit coupling.
\par
Once it is recognized that strong electron-phonon coupling effects can play an
important role in the structural instabilities in the insulating perovskites,
it
then becomes plausible to assume that they can also drive a competition between
structural and superconducting instabilities, when the perovskites are doped.
Indeed, undoped SrTiO$_3$ has an antiferrodistortive instability associated
with
the tilting of the TiO$_6$ octahedra, which is closely analogous to the LTO and
low temperature tetragonal (LTT) phases of LSCO, whereas doped SrTiO$_3$ is
superconducting -- one of the first known members of the family of high-T$_c$
perovskites.  A similar competition between tilt-mode instabilities and
superconductivity is found in BaK$_{1-x}$Bi$_x$O$_3$\cite{Brad}.
Interestingly,
this is accompanied by softening of the 1D O-O bond stretching modes, which
also
play an important role in the cuprates, as will be demonstrated below.

\section{Structural Phase Transitions}

\subsection{Phase Transitions and Pseudogaps}

In LSCO/LBCO and related compounds, a number of structural phase transitions
are observed.  From high temperature to low, the principal phases observed are
a high-temperature tetragonal (HTT) phase, where the octahedra appear to be
untilted, a low-temperature orthorhombic (LTO) phase, in which the average tilt
is about an axis at 45$^o$ to the Cu-O bond, and a low-temperature tetragonal
(LTT) phase, where the tilt axis is along the Cu-O bond.  By substituting Nd
for
La, it is possible to generate an intermediate (Pccn) phase in which the tilt
axis is at an angle between 0 and 45$^o$ to the Cu-O bond.  The transition
temperatures are a strong function of doping, with the LTO transition
temperature monotonically decreasing with hole doping, while a full LTT
transition is found only in LBCO near a doping x=0.125, or in Nd substituted
LSCO.
\par
I have suggested that these transitions are representative of a dynamic
Jahn-Teller (JT) transition.  The LTT
phase would involve a predominantly static distortion, the LTO a dynamic
tunneling between two adjacent LTT tilts, and the HTT tunneling among all four
tilts.  The Pccn phase could be explained as an intermediate phase, where there
is some LTO tunneling, but still a static LTT component.  The HTT$\rightarrow
$LTO transition is a transition between two dynamic JT phases, driven by
intercell coupling associated with corner sharing of planar O's and long range
strain forces.  If the microscopic average local tilt is unchanged between the
LTO and HTO phases, there may be little change in the electronic properties at
the transition -- this appears to be the case experimentally.
However, it is possible that there is an additional
transition {\it within the HTT phase}, associated with the onset of a tilting
instability, which does couple more strongly to electronic degrees of freedom.
Such a crossover from the tunneling phase to an undistorted phase is a common
feature of many dynamic JT transitions\cite{Ham}.
\par
This dynamic tilting onset can be identified with the `pseudogap' phase
recently
found by Hwang, et al.\cite{Hwa} in LSCO, at a temperature $T^*$, Fig. 1.
Below this temperature, there is evidence for significant changes in the
electronic properties of the material: the susceptibility starts to drop
below $T^*$, and the anomalous temperature-dependence of the Hall effect first
appears.  I have suggested that this anomalous temperature dependence follows
from the gradual opening of the short-range-ordered dynamic-JT phase structural
gap, and signals the gradual crossover from the large Fermi surfaces to the
small pockets seen in transport measurements.  Hwang, et al., found a
striking scaling of the pseudogap properties -- the curves at
different dopings all scaled to a universal behavior when plotted as functions
of $T/T^*$.  The LTO phase transition has a similar scaling, with $T_{LTO}
\simeq T^*/2$ for all dopings.
\par
The dynamic JT model is consistent with frozen phonon calculations by Cohen,
et al., which find that in LBCO the HTT phase is unstable, and the LTO phase
metastable with respect to the LTT phase\cite{Pick}.
A similar pseudogap is found in YBCO, associated with peaks in the
susceptibility\cite{R-M} and heat capacity\cite{Lor}.  This is associated with
the onset of a pyramidal tilting instability seen in neutron
diffraction\cite{Schw} and related phonon anomalies\cite{Card,Rama,Kald}. This
was originally interpreted as a spin gap\cite{MP}, possibly with spin-phonon
coupling\cite{Card,NKF},  but the recent evidence is suggestive of a gap in
both
charge and spin channels.
\par
Hence, I would identify this pseudogap phase with the onset of the JT-tilting
instability.  Thus, the calculations of this paper should refer to the $T^*$
transition.  Since, however, most of the experiments are on the LTO
phase, I will make use of the scaling found by Huang, et al. to compare the
calculations for $T^*$ with experimental data for $T_{LTO}\simeq T^*/2$.
\par
It should not be overlooked that such a dynamic JT effect is itself a
potentially important (and highly anomalous) source of low energy excitations,
which could act as additional bosons for causing electrons to pair up.  As
early as 1941, Teller was speculating that the dynamic JT effect could provide
the driving mechanism for (conventional) superconductivity\cite{Tell}.

\subsection{Doping Dependence of LTO Transition}

\par
In the ionic picture for the cuprates, the structural instabilities are driven
by interlayer bond-length mismatch\cite{BGood}.
The unit cell in the CuO$_2$ layer is larger than that in the LaO layer, with
the degree of mismatch measured by the tolerance factor
$$t=d_{LaO}/\sqrt{2}d_{CuO_2},\eqno(1)$$
where $d_{LaO}$ is the effective LaO distance in the LaO plane, with a
similar definition for $d_{CuO_2}$.  Commensurability strains will arise in
LSCO
whenever $t$ is different from unity.  Mismatch problems become more severe at
lower temperatures, gradually causing the CuO$_2$ layers to buckle -- either by
octahedral tilts in LSCO or by layer dimpling in YBCO (there may also be a
pyramidal tilt instability in YBCO).  Indeed, the structural instability can
be interpreted in a purely ionic model, with the transition temperature
proportional to the degree of mismatch.
\par
In undoped La$_2$CuO$_4$, the LTO transition can equally well be described by
either a purely ionic model or a model with considerable electron-phonon
coupling\cite{RM8B} (due to the strong correlation effects, this would be a
spin-Peierls coupling, although the distinction is not obvious at the
mean-field
level).  However, the transition could not be interpreted purely in terms of
Fermi surface nesting; the `bare' phonon frequency $\omega_0$ was always found
to be imaginary, stabilized by anharmonic effects.  This imaginary frequency
could be caused by a combination of covalency effects and interlayer strains,
$$\omega_0^2=\tilde\omega_0^2-\tilde\alpha n_h-\Gamma^*|T_+|\eqno(2)$$
(see Eq. 32c of Ref. \cite{RM8B}).  Here $n_h=1+x$ is the number of holes,
$T_+$ is the interlayer strain, proportional to the lattice mismatch, $\tilde
\omega_0$ is the bare frequency, and $\Gamma^*$ and $\tilde\alpha$ are
appropriate coupling constants.  The middle term of Eq. 2 is a
covalency effect.  Note that there is no electron-hole symmetry within the
antibonding band: the maximum frequency reduction corresponds to an empty band,
$n_h=2$.  The electron-hole symmetry is restored only when both antibonding and
bonding bands are considered.  Thus, the maximum covalent effects arise {\it in
the insulating state}, in a hybridization gap between two atomic levels.  The
last term would indicate a strong phonon-phonon (anharmonic) contribution to
the
phase transitions.
\par
In Appendix A, this analysis is extended to include the doping dependence
of the LTO transition.  Doping reduces the interlayer mismatch in two ways, by
substituting the larger Sr$^{2+}$ for La$^{3+}$, and by hole doping the CuO$_2$
planes.  Insofar as the holes go predominantly onto O's, this amounts to
replacing O$^{2-}$ by the much smaller O$^{1-}$.  Thus, in the doped material
the layer mismatch is reduced, so the tilting instability is driven to lower
temperatures.  Detailed calculations based on this picture can provide a
semiquantitative description of the LTO transition, Fig. 2.  The data in Fig. 2
come from experiments on pure Sr doping (summarized by Takahashi, et
al.\cite{Tak}), and on Nd substitution studies\cite{Buch}.  Once again,
it is found that the transition can be explained equally well by either a
purely
ionic model, or a mixed ionic, electron-phonon model.  The three frames of Fig.
2 correspond to the three models of Fig. 3 of Ref. \cite{RM8B}, with zero (a),
weak (b), or strong (c) electron-phonon coupling.  As in that reference, all
three models provide comparable fits to the data.
\par
In principle, the Nd substitution studies could help clarify the situation,
since Nd is isovalent with La, and Nd substitution should lead to only ionic
size effects, with no doping of the CuO$_2$ layers.  However, even this data
can
be fit by any of the above models.
\par
What is one to make of this?  The principle of Occam's razor would say that
since the purely ionic model (Fig. 2a) provides the simplest interpretation of
the data, then {\it in the absence of any other evidence for electron-phonon
coupling}, this ionic model is preferred.  On the other hand, the cuprates
are high-T$_c$ superconductors, with considerable evidence for phonon
anomalies in the vicinity of T$_c$, and with a strong isotope effect when the
materials are doped away from optimum $T_c$.  In this case, Fig. 2c can be
taken as evidence that the structural transitions are compatible with a strong
electron-phonon coupling.  In fact, much more can be said.  As the next
Section will explain, a careful analysis of {\it the purely ionic model}
strongly suggests that there should be strong electron-phonon coupling.

\section{Cu-O Compressibility and Strong Electron-phonon Coupling}

\subsection{Doping Dependence of the Lattice Mismatch}

\par
In this Section, the ionic model is analyzed in detail, to better understand
the doping dependence of the tolerance factor.  We can begin by relating the
interlayer strain $T_+$, Eq. 2, to the lattice mismatch.  (The middle term of
Eq. 2 has a weak doping dependence, since $n_h$ only varies from 1 to 1.2; it
can be incorporated into a renormalized $\tilde\omega_0$.)  If the equilibrium
cell size of the CuO$_2$ plane is $a_1$ and of the LaO plane $a_2$, then the
two
layers will accomodate to one another at an intermediate cell size $a$, with $a
_2<a<a_1$.  This will involve a compressional strain $e_1\simeq (a-a_1)/a_1$ on
the CuO$_2$ plane, and a tension $e_2\simeq (a-a_2)/a_2$ on the LaO plane.  The
equillibrium lattice size $a$ is found by minimizing the elastic energy
$$E_e={1\over 2}C_1e_1^2+{1\over 2}2C_2e_2^2,\eqno(3a)$$
or
$$a={C_1a_1+2C_2a_2\over C_1+2C_2}.\eqno(3b)$$
(The 2 in front of $C_2$ comes from the fact that there are two LaO layers.)
This ensures that the stresses in the various layers are balanced,
$$T_1=-2T_2,$$
where $T_i=C_ie_i$.  Because of the stress-balance equation, the only free
parameter is $T_2$, or equivalently $a_2-a$. Hence, $T_{LTO}$ is a function
only
of $a_2-a$.  This is convenient, because $a$ is the measured radius, and the
doping dependence of $a_2$ can be estimated from tables of ionic radii.  In
contrast, evaluation of $a_1$ requires knowledge of how hole doping modifies
the
Cu and O radii.  In Appendix B, the parameter $a-a_2$ is estimated as a
function
of doping, and this result is used to derive the theoretical curves in Fig. 2
(see Eq. A2).  As a result of the calculation, the largest contribution to
the doping dependence of the transition is found to be the large ionic size
change associated with O$^{2-}\rightarrow$O$^{-}$, with estimates for the O$^-$
radius that are about 14-21\% smaller than that of O$^{2-}$.
\par
The same calculation can also explain the Nd substitution results.  The Nd ion
is smaller than the La ($1.163\AA$ for Nd$^{3+}$ vs $1.216\AA$ for La$^{3+}$),
so Nd substitution will worsen the mismatch, and hence should enhance the
structural transition temperature.  This is indeed observed: there is a strong
variation of $T_{LTO}$ with Nd content (Fig. 2), which is quantitatively well
described by the present model.  However, the in-plane area of the orthorhombic
cell, $ab$, is virtually unchanged by Nd substitution.  This suggests that
the cell size is fixed by the CuO$_2$ planes, while changing the La-Nd size
leads predominantly to changes in incommensurability strains.  Correspondingly,
it is found that increases in Sr doping lead to large decreases in $ab$.
\par
This can be understood from Eq. 3b.
If $C_1>>C_2$, then $a\simeq a_1$, so the cell area will be independent of the
Nd substitutions in the LaO layer, as observed.  This is consistent with the
expectation that the LaO layer should be more compressible than the CuO$_2$
layer\cite{ScKl}.  Moreover, the
CuO$_2$ layer is already compressed via the distortion of the CuO$_6$
octahedra,
with the four planar O's moving closer to the Cu, the two apical O's moving
away.  (This is a molecular JT effect, which splits the degeneracy of the
$d_{x^2-y^2}$ and $d_{z^2}$ orbitals on the Cu.)  Hence, it will be hard to
produce any additional CuO$_2$ compression (in the absence of hole doping).
\par
In order to explain the pressure dependence of the LTO transition temperature,
$dT_{LTO}/dp<0$, Zhou, et al.\cite{ZCG} suggested that the Cu-O bond length is
more compressible than the La-O bond length.  This result is surprising, and
is not consistent with the above finding, $C_1>>C_2$.  Schilling and
Klotz\cite{ScKl} have suggested that the anomalous $T_{LTO}(p)$ may be due
to pressure induced charge transfer.  While the sign of the charge transfer is
the same as that found in other cuprates, direct Hall effect measurements have
not found evidence for a pressure dependence of the hole concentration in LSCO.

\subsection{Large Electron-phonon Coupling to In-Plane Phonons}

\par
The above calculations reveal that {\it the purely ionic model is not
self-consistent.}  The dominant factor in explaining the Sr doping dependence
of
the LTO transition is the large dependence of the O radius on valence.
However,
it is well known in rare earth compounds, that such large ionic contractions
lead to {\it large electron-phonon coupling}, accompanied by lattice softening
and elastic anomalies\cite{Hews}.
\par
When electron-phonon coupling arises through the dependence of ionic size on
valency, a large number of phonon modes can couple to the distortion.  This
occurs in other mixed-valence systems as well (see, e.g., Mook, et
al.\cite{Hews}).  In the present case, Figure 3 illustrates some of the CuO$_2$
plane phonon modes which should have a strong coupling to the valence
fluctuation.  Thus, if the O$^-$ is localized on a single site, the local
contraction will involve a Cu-Cu bond stretching mode, Fig. 3a.  If it can hop
between adjacent O's across an intervening Cu, it will couple to the O-O bond
stretching mode, Fig. 3b.  These modes are expected to be longitudinal within a
single Cu-O-Cu row on a given plane, but in successive rows, the stretch
distortions can be either in-phase, as in Fig. 3a,b, or out of phase, as in
Fig.
3c,d.  It turns out that, due to vHs coupling, the gaps are larger for the out
of phase modes, Fig. 3c,d.
\par
For completeness, certain other modes are illustrated in Fig. 3e,f.  Figure 3e
shows a related, ferroelectric mode, a 2D version of the mode introduced by
Egami, et al.\cite{Eg}.  In addition to the optical modes, there can also be
coupling to elastic strains (to account for the average, homogeneous effects of
volume reductions) and acoustic phonons (for the local, inhomogeneous effects),
Fig. 3f.  These shear strains function as secondary order parameters for the
LTO
transition.
\par
Since these are all in-plane modes, it is not clear why the actual observed
soft
mode involves out-of-plane tilting.  A possible explanation is developed in
Fig.
3g.  The extra hole is assumed to be partially delocalized in the
$x$-direction,
giving rise to a local shear, shrinking the lattice $x$-axis and increasing the
$y$-axis.  Since the layer is already under compression, the octahedra rotate
about the $x$-axis, to tip the long-bond out of the $a-b$ plane.  This would
explain the highly anomalous feature of the LTT phase, that the Cu-Cu distance
is {\it shorter} along the untilted bond than along the tilted one.  An
analogous effect occurs in the LTO phase.
\par
Recently, Zhou, et al.\cite{ZBG} have also come to the conclusion that in-plane
phonon modes play a large role in the dynamic JT effect in LSCO.

\par
Thus, in analyzing the soft-mode instabilities, the following hypothesis will
be
adopted.  The primary electron-phonon interaction involves a planar distortion,
but as this softens, it interacts with the lower lying LTT-type tilt mode,
leading to an actual soft mode of mixed symmetry.  For a first analysis, this
additional complication of phonon mode-mode coupling will be ignored, and only
the planar mode will be included in the analysis of electron-phonon coupling.

\subsection{Other Problems with the Ionic Model}

There are a number of other features of this structural transition which are
difficult to understand in a simple ionic picture.  First, if the driving
force for the transition is a layer mismatch, with the LaO lattice constant
being too small, it is somewhat surprising that the LaO plane is buckled rather
than flat.  Billenge and Egami\cite{BEg} also questioned the ionic model,
noting
that the CuO$_2$ planes in Nd$_{2-x}$Ce$_x$CuO$_4$ showed local buckling, even
though in the ionic model, they are supposed to be under tension.
\par
More importantly, Fig. 4 shows the doping dependence of the Cu-O
bond length found by Radaelli, et al\cite{Rad}.  The detailed form of this
dependence is inconsistent with a purely ionic model.  There are actually a
number of problems. One problem lies in the temperature dependence at fixed
$x$.
If the tolerance factor were the driving force for the transition, it would be
expected that the Cu-O bond length would act like an order parameter, having a
strong temperature dependence in the LTO phase.  Instead, this bond length has
a
strong $x$ dependence, but is nearly independent of $T$ in the LTO phase,
whereas it has a strong $T$ dependence and weak $x$ dependence in the HTT
phase.
The LTO-HTT transition occurs when these two bond lengths are equal.
\par
This leads to a second problem: {\it on each side of the phase transition,
the stable phase is the one with the longer Cu-O bond length}!  This is
opposite to the prediction of the ionic model, which says that the driving
force
for the transition is to reduce the Cu-O bond length.
\par
There is also another potential problem.  The dot-dashed line in Fig. 4 shows
the doping dependence of the equilibrium Cu-O bond length, which agrees with
the
measured value in the LTO phase, but {\it changes much more rapidly with doping
than the experimental value in the HTT phase}.  The break in slope cannot be
explained by the tolerance factor passing through the value unity.
Theoretically, this is not expected to happen until $x\simeq 0.42$.  More
importantly, the break in slope occurs at different $x$ values at different
temperatures.  However, the HTT data all come from the overdoped regime, $x>
0.15$.  It is known\cite{Rad} that the uniform phase is metastable against
phase separation in this regime, and moreover, that the doping holes go into
additional orbitals -- in particular, the Cu $d_{z^2}$ orbitals.  Hence, the
change in slope may be a signature of this {\it electronic} effect --
experiments at higher temperatures and lower doping are clearly desirable.
Again, this shows that electronic effects must be included in order to
understand the structural phase transitions.
\par
The dynamic JT model can provide at least a partial explanation of these facts.
It is important to recall that the onset of the tilting instability
actually occurs at $T^*$.  Hence, since both the LTO and HTT phases are dynamic
JT phases, the HTT-LTO crossover will be controlled by long-range strain
forces,
and need not involve a substantial change in the Cu-O bond length.

\section{Linear Electron-Phonon Coupling and the O-O Bond Stretching Mode}

\par
Figure 3 illustrates some of the phonon modes which could couple to the large
O-atom valence fluctuations.  To better understand the nature of the resulting
electron-phonon coupling, the modification of the electronic band structure
can be calculated, in the presence of a static distortion of the appropriate
symmetry.  It is assumed that the distorted bond lengths lead to a modulation
of
the Cu-O hopping patameter $t_{CuO}$, $t_{CuO}\rightarrow t_{CuO}\pm\delta t$.
Assuming the standard three-band model for the CuO$_2$ planes, Figure 5
illustrates the resulting dispersions, showing that several modes can open a
gap
at the Fermi level.  The calculations are given in Appendix C.  (The figure is
calculated assuming bare band parameters, $t=1eV$, $\Delta =4eV$; for the
renormalized bands\cite{RMX}, the energy scale should be reduced by a factor of
$\sim 4$.)
\par
{}From Fig. 5, it can be seen that the largest gap is associated with the O-O
bond stretching mode, Fig. 5d.  The Cu-Cu bond stretching mode has a comparable
gap, but with a strong $\vec k$-dependence, and vanishes at the vHs (5c).  This
difference is discussed in Appendix C.  No gap is opened for the FE mode (5e)
or
the strains (5f), although in the latter case, the anisotropy splits the vHs
degeneracy along the $X$ and $Y$ axes.
The differences between the $q_1$ nesting (Fig. 5a,b) and $q_0$ nesting,
(Fig. 5c,d) lie mainly in the different zone folding associated with the
doubling of the unit cell.  For $q_1$ nesting, the $X$ point is mapped into
$\Gamma$, while $M$ is mapped into $Y$; on the other hand, $q_0$ maps $M$ into
$\Gamma$ and $X$ into $Y$.  For ease in comparison, Figs. 5e,f are illustrated
including a $q_0$-type zone folding.
\par
The O-O bond stretching mode looks like an ordinary 1D Peierls instability cum
CDW, and has the form postulated for the vHs-JT effect.  Tang and
Hirsch\cite{TH} had analyzed the CDW transition
several years ago, and found that it did not couple to the vHs's.  However,
they
used a Cu-only model, so their CDW corresponds to the Cu-Cu bond stretching
mode, which indeed does not split the vHs degeneracy, Fig. 5c.  From Eq. C9, it
can be seen that the O-O bond stretching mode, Fig. 3d and the shear strains,
Fig. 3f, involve a {\it linear} electron-phonon coupling, as opposed to the
quadratic coupling to the tilt-mode phonons.  This linear coupling will lead to
large {\it polaronic} effects, since the more localized the hole is, the closer
to O$^-$ is the corresponding oxygen, with a much larger local contraction.
Such polaronic effects have often been postulated to play a role in the high-T
superconductivity.
\par
Direct evidence for anomalously strong electron-phon\-on coupling of the O-O
bond stretching mode has been found by neutron scattering measurements, both in
LSCO and in YBCO, and closely analogous results are found in Ba$_{1-x}$K$_x
$BiO$_3$ (BKBO).  In undoped LSCO, this one-dimensional bond stretching mode is
found near 20THz, with a weak softening near the $X$-point.  Upon doping, the
softening substantially increases\cite{PiRe}.  A similar result is found in
YBCO\cite{PiRe}.
\par
In BKBO, it has generally been believed that the structural anomalies are
associated with three dimensional breathing modes.  However, in the heavily
doped materials there are soft modes associated with BiO$_6$ octahedral tilts,
similar to those in LSCO\cite{Brad}.  Moreover, no anomalous softening of the
three-dimensional breathing modes is found.  Instead, the one-dimensional
$Bi-O$ bond stretching mode softens, again in complete analogy to the
cuprates\cite{Brad}.  It has proven difficult to model this softening in a
lattice dynamical model, assuming the local symmetry is cubic.  These results
suggest that in BKBO the structural and superconducting anomalies may be
associated with a 3D vHs, as in the Bilbro-McMillan model\cite{BilM}, which has
been applied to the A15's and BaPb$_{1-x}$Bi$_x$O$_3$\cite{BPBO}.

\section{Polaronic Hamiltonian}

\par
Let us briefly summarize the above discussion.  The analysis of the doping
dependence of $T_{LTO}$ shows that, while interlayer mismatch is a dominant
driving force, a purely ionic model cannot provide an adequate description of
the data.  Moreover, even within the ionic model, most of the accomodation of
the mismatch with doping is found to be due to a large variation of the
in-plane
Cu-O bond length with doping (chemically, with the valence change
O$^{2-}\rightarrow$O$^-$).  This bond length change itself produces a large
electron-phonon coupling, with a number of in-plane modes.  The strength of
coupling can be estimated from the size of the gap resulting from a CDW
distortion: the largest gap is associated with the O-O bond stretching mode,
Fig. 5d, due to the splitting of the vHs degeneracy.  However, the symmetry of
this distortion is not that of the LTO phase, so this distortion can only be
present in the form of a local, short-range order, possibly dynamic.  Such
short
range order has already been postulated in terms of the tilting
modes\cite{RM8C}.  This picture is supported by experiments, which find
significant softening of the relevant modes, and LDA frozen phonon
calculations,
which find a significant phonon softening for this mode.
\par
These results suggest an important modification to the earlier
calculations of the LTT-LTO-HTT transitions as a series of dynamic vHs-JT
effects\cite{RM8A,RM8B,RM8C,RMX}.  In these papers, the soft mode was
identified
with the experimentally observed mode, associated with octahedral tilt (plus an
associated shear strain).  This led to an unusual electron-phonon coupling,
quadratic in the tilt angle $\theta$, and hence in the phonon
creation operator -- in violation of Migdal's theorem.  Nevertheless, the
resulting coupling is estimated to be quite large\cite{RMX,SongAn}.
The present results suggest that the tilting is a secondary effect, and the
primary driving force is associated with in-plane modes (particularly the O-O
bond stretching mode) that have a linear electron-phonon coupling.  Hence,
there
are important {\it polaronic} contributions, which will be estimated below.
Note that many of the predictions of the model are not sensitive to the
particular phonon mode; these include the occurence of a pseudogap\cite{RM8A},
and of a dynamic JT effect\cite{RM8C}, which has apparently been
observed experimentally\cite{Hamm,Bill}.
\par
In this section it will be assumed that the principal electron-phonon coupling
is with the planar bond stretching mode, and that the tilt mode softens first
due to phonon mode-mode coupling with these planar modes.  The complications
due
to this mode-mode coupling will here be ignored, leaving a problem with a
linear
electron-phonon coupling, which can be analyzed using standard techniques.
It should be noted that the resulting model is essentially a vHs-driven CDW
instability, and hence very similar to a number of earlier
calculations\cite{TH,SSB,All}.  The differences are in the identification of
the dominant phonon mode (Fig. 5d), and the realization that the physics is
dominated by dynamic, short-range fluctuations, having a different
symmetry from the macroscopic average.

\subsection{Starting Hamiltonian}

The three-band Hamiltonian can be written:
$$H=H_e+H_{ph}+H_{e-ph},\eqno(4)$$
where the electronic Hamiltonian is
$$H_e=\sum_j\bigl(\epsilon_d d^{\dagger}_jd_j+
\sum_{\hat\delta}\epsilon_p p^{\dagger}_{j+\hat\delta}p_{j+\hat\delta}
+\sum_{\hat\delta}t[d^{\dagger}_jp_{j+\hat\delta}+(c.c.)]$$
$$+\sum_{\hat\delta^{\prime}}t_{OO}[p^{\dagger}_{j+\hat\delta}p_{j+\hat\delta^
{\prime}}+(c.c.)]
+Un_{j\uparrow}n_{j\downarrow}\bigr),\eqno(5)$$
where $d^{\dagger}$ ($p^{\dagger}$) is a creation operator for holes on Cu (O),
j is summed over lattice sites, $\hat\delta$ over nearest neighbors,
$\hat\delta
^{\prime}$ over next-nearest (O-O) neighbors, and c.c. stands for complex
conjugate.  $\Delta=\epsilon_d-\epsilon_p$ is the Cu-O splitting, $t$ ($t_
{OO}$) is the Cu-O (O-O) hopping energy, and $U$ is the on-site Coulomb
repulsion.  In a slave boson calculation, the Coulomb term does not appear
explicitly, but via correlation effects, which renormalize the one-electron
band parameters.  In this case, the eigenenergy of the one-electron
Hamiltonian,
becomes
$$E_{\vec k}={\Delta\over 2}+\sqrt{({\Delta\over 2})^2+4t^2(s_x^2+s_y^2)},
\eqno(6)$$
where $\Delta =\epsilon_d-\epsilon_p$, and I have introduced the notation
(Appendix C) $s_i=sin(k_ia/2)$, $c_i=cos(k_ia/2)$.
\par
The pure phonon term is
$$H_{ph}={1\over 2}\sum P(\lambda ,\vec q)P(\lambda ,-\vec q)$$$$
+{1\over 2}\sum
\omega_0^2(\lambda ,\vec q)Q(\lambda ,\vec q)Q(\lambda ,-\vec q)\eqno(7)$$
where $Q(\lambda ,\vec q)$ ($P(\lambda ,\vec q)$) is the phonon position
(momentum) operator.  The form of $H_{e-ph}$ is discussed below. In this paper,
only a highly simplified model will be analyzed, to
illustrate the nature of the polaronic coupling.  Among the simplifications
assumed are (1) $t_{OO}=0$; (2) correlation effects are small enough that they
can be incorporated by the use of renormalized band parameters; (3) the only
relevant phonon mode is the 1D O-O bond stretching mode.
\par
Technically, assumptions (1) and (2) are incompatible, but this is unimportant
for the present, illustrative purposes.  Assumption (1) forces the vHs's to
fall
at half filling, where correlation effects are large, and {\it suppress} the
electron-phonon coupling, particularly near the insulating phase\cite{KiT,RMX}.
In this case, spin-Peierls coupling should be included, and the resulting
analysis is not greatly modified.  This is the topic of a future
publication\cite{RMXC}.  In the doped material, correlation effects are
relatively small, and the present analysis should apply.  However, a non-zero
$t_{OO}$ is required to move the vHs into this regime, and the extra couplings
introduced by this term greatly complicate the analysis, without significantly
affecting the underlying physics.
\par
Assumption (3) restricts the analysis to a single branch of phonons, which
includes the O-O bond stretching mode.  The distinction between the
phonons of the form of Fig. 4b and 4d lies in the direction of the phonon
propagation vector, corresponding to two different cuts along this branch. For
the cut of Fig. 4b, $\vec q=\zeta\vec q_1$, with $\vec q_1=(\pi /a,0)$ and
$\zeta\in [0,1]$; for cut 4d, $\vec q=\zeta\vec q_0$, $\vec q_0=(\pi /a,\pi /
a)$.  Only the latter cut will be analyzed in detail.  The corresponding phonon
operators will be denoted so $Q_{\vec q}=Q(\lambda ,\vec q)$, $P_{\vec q}=P(
\lambda ,\vec q)$, and $\omega_{\vec q}=\omega_0(\lambda ,\vec q)$.  It should
be noted that this same phonon, at $\vec q=0$, has the character of a FE mode.
In a more detailed treatment of the phonons, this should correspond to a
nearly pure O-O vibration, with the xO's and yO's vibrating out of phase.
\par
The analysis of Section III (Figs. 4,5) suggests that the electron-phonon
coupling should be included via the dependence of $t$ on the ionic
separation\cite{Pyt}
$$t_{ij}=t+(\vec u_j-\vec u_i)\cdot\vec\nabla t_{ij},\eqno(8a)$$
where the $\vec u_i$'s are the displacements of the $i$th atoms.
The $\vec u_i$ can be expanded in terms of the phonon normal mode coordinates
as
$$\vec u_i=\sum_{\lambda, \vec q}{\vec e(\lambda ,\vec q)\over (mN)^{1/2}}
e^{i\vec q\cdot\vec R_i}Q(\lambda ,\vec q),\eqno(8b)$$
with $\vec e(\lambda ,\vec q)$ the polarization vector and $m$ the ionic mass.
In the present, single branch approximation, defining $g_{\vec q}=-\vec e(\vec
q,\lambda)\cdot\vec\nabla t/\sqrt{mN}$, the
electron-phonon coupling with the O-O bond stretching mode becomes
$$H_{e-ph}=\sum_{\vec k,\vec q}Q_{\vec q}\tilde g_{\vec q}(\vec k)(d^{\dagger}_
{\vec k}p_{x,\vec k-\vec q}+p^{\dagger}_{x,\vec k-\vec q}d_{\vec k})
,\eqno(9)$$
with
$$\tilde g_{\vec q}(\vec k)=2g_{\vec q}c_x.\eqno(10)$$
(To describe the Cu-Cu hopping, it is only necessary to make the
substitution $c_x\rightarrow -cos(k_x-q_x)a/2$ in Eq. 10.)
\par
The resulting Hamiltonian, Eqs. 4-7,9, is almost exactly of the form analyzed
in Ref. \cite{RM8A} (after Balseiro and Falicov\cite{BFal}) to study the
competition between Peierls instability and superconductivity.  The differences
are (1) the Fermi surface is square at half filling, since the O-O
hopping term has been neglected (this was done only for simplicity, and an
appropriate $t_{OO}$ may readily be reintroduced); and (2) the
electron-phonon coupling $\tilde g(\vec k)$ now has an important wave number
dependence.

\subsection{Link Operator Formalism}

The electron-phonon coupling term can be eliminated by a canonical
transformation\cite{Pyt,BP,Hols,Mahan}:
$$\bar H=e^{-iS/\hbar}He^{iS/\hbar}\simeq H+{i\over\hbar}[H,S].\eqno(11)$$
In the usual small polaron theory (e.g., Ref.~\cite{Mahan}), this
transformation
must be carried out to all orders in $S$ to correctly describe the polaronic
band narrowing.  However, in the present problem, the function $S$ is quite
complicated, due to the 2D nature of the problem.  Hence, in what follows, I
introduce the following approximation scheme.  I first find an $S$ which
exactly
eliminates the electron-phonon coupling in first order, Eq. 11.  I then show
that in the transformed Hamiltonian $\bar H$, the hopping parameter is
renormalized to $t\rightarrow t(1-{\cal S})$.  This is precisely the form of
the lowest order band narrowing found in conventional small polaron theory, and
${\cal S}$ can be shown to have exactly the same interpretation.  Hence, it
will
be {\it assumed} that the role of higher order terms is essentially the same in
the present model, to transform the renormalization to $(1-{\cal S})\rightarrow
e^{-{\cal S}}$.  The results are given in Section V.C.4.
\par
$S$ can be expressed in terms of the (generalized) link operators as
$$S=i\sum_{\vec k,\vec q}(Q_{\vec q}A_{\vec k,\vec q}-iP_{\vec q}B_{\vec k,\vec
q}),\eqno(12)$$
where
$$A_{\vec k,\vec q}=\sum_{i=1,9}f_{i\vec k,\vec q}\rho_{i-,\vec k,\vec
q},\eqno(13a)$$
$$B_{\vec k,\vec q}=\sum_{i=1,9}h_{i\vec k,\vec q}\rho_{i+,\vec k,\vec
q},\eqno(13b)$$
$$\rho_{1\pm ,\vec k,\vec q}=d_{\vec k}^{\dagger}p_{x\vec k-\vec q}
\pm p_{x\vec k-\vec q}^{\dagger}d_{\vec k}
,\eqno(14a)$$
$$\rho_{2\pm ,\vec k,\vec q}=p_{x\vec k}^{\dagger}d_{\vec k-\vec q}
\pm d_{\vec k-\vec q}^{\dagger}p_{x\vec k}
,\eqno(14b)$$
$$\rho_{3\pm ,\vec k,\vec q}=d_{\vec k}^{\dagger}p_{y\vec k-\vec q}
\pm p_{y\vec k-\vec q}^{\dagger}d_{\vec k}
,\eqno(14c)$$
$$\rho_{4\pm ,\vec k,\vec q}=p_{y\vec k}^{\dagger}d_{\vec k-\vec q}
\pm d_{\vec k-\vec q}^{\dagger}p_{y\vec k}
,\eqno(14d)$$
$$\rho_{5\pm ,\vec k,\vec q}=d_{\vec k}^{\dagger}d_{\vec k-\vec q}
\mp d_{\vec k-\vec q}^{\dagger}d_{\vec k}
,\eqno(14e)$$
$$\rho_{6\pm ,\vec k,\vec q}=p_{x\vec k}^{\dagger}p_{x\vec k-\vec q}
\mp p_{x\vec k-\vec q}^{\dagger}p_{x\vec k}
,\eqno(14f)$$
$$\rho_{7\pm ,\vec k,\vec q}=p_{y\vec k}^{\dagger}p_{y\vec k-\vec q}
\mp p_{y\vec k-\vec q}^{\dagger}p_{y\vec k}
,\eqno(14g)$$
$$\rho_{8\pm ,\vec k,\vec q}=p_{x\vec k}^{\dagger}p_{y\vec k-\vec q}
\mp p_{y\vec k-\vec q}^{\dagger}p_{x\vec k}
,\eqno(14h)$$
$$\rho_{9\pm ,\vec k,\vec q}=p_{y\vec k}^{\dagger}p_{x\vec k-\vec q}
\mp p_{x\vec k-\vec q}^{\dagger}p_{y\vec k}
.\eqno(14i)$$
These link operators are the generalization of the operators introduced by
Halperin and Rice\cite{HaRi} into the problem of density wave instabilities of
an electron gas, and first applied to the cuprates by Schulz\cite{Sch2}.  They
also bear a close resemblance to the valence bond operators studied by Affleck
and Marsden\cite{KAM}.
\par
A few technical points about these operators should be noted.  First, the
Hamiltonian, Eq. 4, can be rewritten in terms of these operators:
$$H_e=\sum_{\vec k}\bigl({\epsilon_d\rho_{5-,\vec k,0}+\epsilon_p(\rho_{6-,\vec
k,0}+\rho_{7-,\vec k,0})\over 2}$$$$
+t_{x\vec k}\rho_{1-,\vec k,0}+t_{y\vec k}\rho_{3-,\vec k,0}\bigr),\eqno(15a)$$
$$H_{e-ph}=\sum_{\vec k,\vec q}Q_{\vec q}\tilde g_{\vec q}(\vec k)\rho_{1+,\vec
k,\vec q},\eqno(15b)$$
with $t_{i\vec k}=2its_{i\vec k}$, $s_{i\vec k}=sin(k_ia/2)$. Secondly, the
sign
change in Eqs. 14e-i arises because the hopping term in $H_e$ is imaginary
($is_x$).  Finally, the set of $\rho$'s is actually overcomplete, since, e.g.,
$\rho_{2\pm ,\vec k,\vec q}=\pm\rho_{1\pm ,\vec k-\vec q,-\vec q}$.
Because of the 2D nature of the problem, finding the $f_i$'s and $h_i$'s
requires inverting an $18\times 18$ matrix.  The solution is given in Appendix
D.

\subsection{Polaronic Effects}

\subsubsection{Effective Interaction}

\par
Once the linear electron-phonon coupling is eliminated, we are left with
a complicated Hamiltonian involving electron-electron and quadratic
electron-phonon interactions, describing a rich variety of phenomena, including
phonon softening, structural instability (CDW formation), and
superconductivity.
In this paper, no attempt is made to sort out all of these effects, but only
to summarize the various phenomena which are expected.  For this purpose, two
simpler, approximate solutions are also presented in Appendix D: one is a
three-band, but quasi-1D solution, in which coupling terms involving only the
$x$ direction are retained.  The second approximate solution is an effective
one-band solution, in which the three-band electronic Hamiltonian is
diagonalized, and only transitions within the antibonding band are retained.
The resulting equations are qualitatively similar to standard results for
polaronic calculations.  However, there are some subtleties, involving band
mixing and a potential $\vec k$ dependence of the couplings.
\par
The qualitative nature of the resulting polaronic Ham\-il\-ton\-ian can be seen
by ignoring all the link operators except $\rho_1$ in Eqs. 12-14, and
approximating
$$f_{1\vec k,\vec q}\simeq{\hbar\tilde g_{\vec q}(\vec k)\beta_{\vec q}\over
\hbar^2\omega_{\vec q}^2-(E_{\vec k}-E_{\vec k-\vec q})^2},\eqno(16a)$$
$$\beta_{\vec q}\simeq{\hbar^2\omega^2_{\vec q}\Delta\over\Delta^2+16t^2}.
\eqno(16b)$$
$$h_{1\vec k,\vec q}\simeq{f_{1\vec k,\vec q}4t^2
\over\Delta\hbar\omega_{\vec q}^2}\eqno(16c)$$
(see Eq. D11).
The commutator of $S$ with $H_{e-ph}$ gives the residual interaction terms,
$${i\over\hbar}[H_{e-ph},S]=H_{e-e}+H_{e-ph}^{(2)},\eqno(17)$$
with (Eq. D9)
$$H_{e-e}=-\sum_{\vec k,\vec k^{\prime}}h_{1\vec k^{\prime},\vec q}
\tilde g_{\vec q}(\vec k)\rho_{1+,\vec k,\vec q}\rho_{1+,\vec k^{\prime},\vec
q}.\eqno(18)$$
The remaining term, $H_{e-ph}^{(2)}$ is complicated in the three-band model,
because it mixes in different link operators, $\rho$ (see Eq. D13).  However,
it simplifies in the one-band model, to
$$H_{e-ph}^{(2)}=-\sum_{\vec k}{f_{1\vec k}Q_q^2\over\hbar}[\alpha_{\vec k+\vec
q}\psi^{\dagger}_{\vec k+2\vec q}\psi_{\vec k}-\alpha_{\vec k-\vec q}
\psi^{\dagger}_{\vec k+\vec q}\psi_{\vec k-\vec q}],\eqno(19)$$
where $\psi$ is the wave function of the antibonding band, $f_1$ (Eq. D20)
is closely related to $f_{1\vec k,\vec q}$, above, and $\alpha_{\vec k}$ is
given by Eq. D18b.

\subsubsection{Polaron Binding Energy}

\par
The polaron binding energy can be defined from Eq. 18 as
$$E_P=lim_{\Delta E\rightarrow 0}\sum_{\vec k}h_{1\vec k,-\vec q}
\tilde g_{\vec q}(\vec k)$$
$$=\sum_{\vec k}{|\tilde g_{\vec q}(\vec k)|^2\eta\over 2\omega_{\vec q}^2}$$
$$={4\beta^2t^2\over ma^2\omega_{\vec q}^2},\eqno(20)$$
where $\Delta E=E_{\vec k}-E_{\vec k-\vec q}$, $\eta =1/2(1+(\Delta /4t)^2)$,
and the electron-phonon coupling constant can be roughly estimated as $|g_{\vec
q}|\sim -\vec\nabla t/\sqrt{mN}=\beta t/(a/2)\sqrt{mN}$, where $\beta$ is a
dimensionless parameter, $\sim 3-3.5$\cite{All,NKF}.  This estimate relies on a
simple tight-binding model.  A better way to estimate $g$ would be via an LDA
frozen phonon calculation\cite{Pick}.
\par
Using numerical estimates of $t=0.25eV$, $\beta =3.5$, and $\hbar\omega_{\vec
q}
\simeq 50meV$, a modest value $E_P\simeq 6meV$ is found.  However, since
$\Delta
E <<\hbar\omega_{\vec q}$, Grilli and Castellani\cite{GC} showed that
correlation effects should not reduce $g$.  This would correspond to using a
bare value $t\simeq 1.3eV$ in the above estimate, leading to $E_P\simeq 170
meV$.

\subsubsection{Phonon Softening}

\par
In a mean field decoupling, $H_{e-ph}^{(2)}$ leads to a phonon mode softening,
$$\tilde\omega_0^2=\omega_0^2+\sum \gamma_{\vec k}
{(n_{\vec k}-n_{\vec k-\vec q_0})(E_{\vec k}-E_{\vec k-\vec q_0})\over
(E_{\vec k}-E_{\vec k-\vec q_0})^2-\hbar^2\omega_0^2},\eqno(21)$$
where $n_{\vec k}=<\psi^{\dagger}_{\vec k}\psi_{\vec k}>$, and the precise form
of $\gamma_{\vec k}$ is not important for present purposes (see Eq. D22).  In a
self-consistent calculation, the bare $\omega_0$ in the denominator of Eq. 21
would be replaced by $\tilde\omega_0$.  Making this substitution, a
phase transition occurs when $\tilde\omega_0$ is renormalized to zero, or
$$\omega_0^2=-\sum \gamma_{\vec k}{n_{\vec k}-n_{\vec k-\vec q_0}
\over E_{\vec k}-E_{\vec k-\vec q_0}}.\eqno(22)$$
\par
In Eq. 22, the $E$'s can be replaced by $E^{\prime}=E-E_F$.  Now if the Fermi
energy falls at the vHs, then $E^{\prime}_{\vec k-\vec q_0}=-E^{\prime}_{\vec
k}
$, so the transition temperature $T_0$ is given by
$$\omega_0^2=<\gamma_{\vec k}>\int_{-B}^B{N^*(E)dE\over 2E}tanh({E\over 2k_B
T_0}),\eqno(23a)$$
where $<\gamma_{\vec k}>$ is the average of $\gamma_{\vec k}$ over angles, $B$
is the half bandwidth, and $N^*(E)$ is an effective density of states (dos)
$$N^*(E)={1\over\pi}\sum_{k_y}{\gamma_{\vec k}\over <\gamma_{\vec k}>|\partial
E/\partial k_x|}.\eqno(23b)$$
The integral in Eq. 23 must be evaluated numerically, but certain general
features can be directly extracted.  The result is sensitive to the angle
dependence of $\gamma_{\vec k}$.  If $\gamma_{\vec k}$ is a constant, then the
results are the same as found earlier\cite{RM8A}: $N^*(E)$ is the actual dos,
which diverges logarithmically at the vHs, greatly enhancing $T_0$.  This
appears to be consistent with the gap found for the O-O bond stretching mode,
Fig. 5d.  However, the approximation of retaining only one pair of link
operators is too crude to reproduce this feature, and the model $\gamma$'s
are found to have a more complicated angular dependence.  Indeed, it is
expected, on analogy with the results of Appendix C, that extreme care must be
taken to derive the correct angle-dependence of the gap.
\par
For other phonon modes, however, $\gamma_{\vec k}$ is expected to have an
important angle dependence.  Thus, for the Cu-Cu bond stretching mode, $\gamma_
{\vec k}\simeq\gamma_0sin^2k_xa$ (see the discussion below Eq. C8).  In this
case, $N^*(E)\simeq 16\sqrt{B^2-E^2}/(\pi B)^2$, remaining finite at the vHs,
$E=0$, and
$$k_BT_0\simeq 1.14{B\over 2}e^{-1/\lambda},\eqno(24a)$$
with
$$\lambda=4<\gamma_{\vec k}>N_1/\omega_0^2,\eqno(24b)$$
with $N_1=16/\pi^2B$.
Thus, for the Cu-Cu bond stretching mode, the CDW has a logarithmic
instability,
and not the $ln^2$ instability expected for a vHs-related singularity, due to
the fact that the CDW gap for this mode vanishes at the vHs, Fig. 5c.  This is
consistent with earlier work\cite{TH}.

\subsubsection{Polaronic Band Narrowing}

\par
The term $H_{e-ph}^{(2)}$ also induces a polaronic reduction of the
bandwidth\cite{Hols}.  In the Appendix, this reduction is calculated to lowest
order in $S$ as $t\rightarrow t(1-{\cal S})$, Eq. D15, with
$${\cal S}\simeq\sum_{\vec q}\bigl(
{\tilde g_{\vec q}(\vec k)h_{1\vec k,-\vec q}\over \hbar\omega_{\vec q}}\bigr)
(N_{\vec q}+{1\over 2}),\eqno(25a)$$
where the Bose function is
$$N_{\vec q}={1\over e^{\hbar\omega_{\vec q}/k_BT}-1}.\eqno(25b)$$
In conventional polaron theory, inclusion of higher order commutators in $S$
leads to an exponential bandwidth reduction.  Hence, I {\it hypothesize} that
the main effect of including higher order commutators with $S$ in Eq. 11 would
be the renormalization
$$t\rightarrow te^{-{\cal S}}.$$
However, since $\cal S$ can have either sign, a more appropriate approximation
would be
$$t\rightarrow {2t\over 1+e^{2{\cal S}}}.\eqno(25c)$$
By comparing Eq. 25a and Eq. 20, it can be seen that at low temperatures ($N_{
\vec q}\simeq 0$) ${\cal S}\simeq E_P/2 \hbar\omega_{\vec q}$, so using the
above estimate for $E_P$, $exp({\cal S})\simeq 5$.
\par
In all of the above expressions, the Fermi energy enters only through the
Fermi-Thomas function, $n_{\vec k}$.  The polaronic band narrowing, Eq. 25,
however, is {\it independent of $n_{\vec k}$, and hence of $E_F$}!  It depends
only on the details of the band structure; in particular, {\it there is a
significant band narrowing only when}
$$\hbar\omega_{ph}>|E_{\vec k+\vec q/2}-E_{\vec k-\vec q/2}|.\eqno(25d)$$
When this condition is satisfied, the bandwidth reduction is similar to the
usual phonon renormalization, $1/(1+\lambda )$, with $\lambda\sim e^{\cal S}$.
\par
The condition, Eq. 25d, is discussed in more detail in the following section.
It is found that {\it this condition can be satisfied only in the immediate
vicinity of the vHs}!  This may explain the presence of `extended
vHs's'\cite{PE2} in the cuprates.  Figure 6 illustrates a calculation of the
modified band structure of Bi-2212, if it is assumed that there is a phonon
renormalization of the energy bands by a factor $\lambda =3$ within
$\omega_{q}=
40meV$ of the Fermi level.  The value $\lambda =3$ is in satisfactory agreement
with the renormalization factor $exp({\cal S})$ found above.  Moreover, a
similar band narrowing is found near the vHs in YBa$_2$Cu$_4$O$_8$\cite{PE1},
{\it even though the vHs is $\sim 19meV$ away from the Fermi level}.
\par
If terms in $t_{OO}$ are retained, the parameters $t$ and $t_{OO}$ are
renormalized by different factors, so polaronic effects could lead to a change
in the shape of the Fermi surface.  Analogous effects found in the f-electron
problem\cite{HewNew} suggest that direct O-O hopping may be more seriously
affected than Cu-O hopping.
\par
In ordinary small polaron theory\cite{Mahan}, the higher order commutators play
two roles: for the {\it coherent hopping}, they lead to a polaronic band
which becomes increasingly narrow as temperature is raised, as in Eq. 25.
However, they also lead to an {\it incoherent} hopping due to nondiagonal
transitions (in which the number of phonons bound to the electron changes
during
the hopping process).  This latter effect is not accounted for in the present
calculation, which treats $S$ to lowest order.

\subsubsection{CDW vs. Superconductivity}

\par
The CDW and superconducting instabilities can also be studied in terms of $H_{
e-e}$.  For the CDW instability, only the term in $\vec q=\vec q_0$ is
retained\cite{BFal,RM8A}.  With this approximation, $H_{e-e}$ becomes, in mean
field,
$$H_{CDW}=-G_0\sum_{\vec k}c_x\rho_{1+,\vec k,\vec q_0}.\eqno(26)$$
Since $H_e+H_{CDW}$ is quadratic, it can be diagonalized by the
Bogoliubov-Valatin transformation.  The resulting dispersion satisfies Eq.
C6, if $G_0=2\delta t$.  Again, the present approximate models lead to a $G_0$
with a more complicated angular dependence.
\par
For superconductivity, those terms in Eq. 18 are retained, for which $\vec k^
{\prime}=-\vec k$, and $\sigma^{\prime}=-\sigma$, where the $\sigma$'s are the
spin indices, which had been neglected up to now.  This leads to a
conventional,
vHs-enhanced superconductivity\cite{RM8A}, except that {\it the gap equation
contains an angular factor similar to $\gamma_{\vec k}$}!  Such an angular
factor can, in principle, lead to a non-s-wave gap.

\section{Discussion}

\subsection{Ionic-Covalent Crossover and the vHs}

In their discussion of electron-phonon coupling induced by valence fluctuations
in f-electron metals, Sherrington and Riseborough (SR)\cite{SheRi} introduced a
Hamiltonian with a somewhat different form of $H_{e-ph}$.  SR assumed a direct
coupling to the valence fluctuation, which in the present context would give
$$H_{e-ph}^{SR}=\sum_{\vec k,\sigma}g_{\vec k}p^{\dagger}_{\vec k+\vec q_0,
\sigma}p_{\vec k,\sigma}Q_0,\eqno(27)$$
Here, while the tight-binding electron-phonon coupling has a
different form, Eq. 9, the resulting Hamiltonian automatically includes a term
like Eq. 27, but with quadratic phonon coupling, Eq. D13.
\par
The present calculation sheds some light on a feature that SR
were attempting to describe: the competition between covalency and ionicity.
They pointed out that earlier theories of electron-phonon coupling in f metals
ranged from strongly covalent\cite{AlL}, with each rare earth ion having
average
values of ionic size and f-electron occupation number, to strongly
ionic\cite{AnCh}, with well-defined $f^n$ and $f^{n-1}$ ions.  SR suggested
that
this competition is represented by terms describing hopping, $t$, and
electron-phonon coupling, $H_{e-ph}$: $H_{e-ph}<<t$ represents the covalent
limit, where the valence can freely fluctuate, while the opposite limit,
$H_{e-ph}>>t$, is the ionic limit, in which each ion has a well defined
valence.
\par
This can be restated: the strong electron-phonon
coupling is associated with valence fluctuations: a hole hops onto an O$^{2-}$,
causing its radius to collapse, leading to an enhanced local value of $t$.  For
the collapse to occur, the phonon must be able to follow the electronic
motion, $\omega_{ph}\sim (hopping\ rate)$.  For an `isolated CuO$_6$ molecule',
this hopping rate is $\sim t/\hbar>>\omega_{ph}$. Near half filling,
strong correlations reduce the hopping, but the inequality still holds,
$J\simeq
130meV>\hbar\omega_{ph}\simeq 30-70meV$.
\par
In a band, however, the electronic motion is restricted by the motion of other
electrons, and the relevant frequency becomes $\vec q\cdot\vec v_F$, where $q$
is the phonon wave number and $v_F$ the Fermi velocity.  At an arbitrary point
on the Fermi surface, this quantity is of comparable magnitude to $t/\hbar
$, but $v_F$ {\it vanishes at a vHs}!  Figures 7a,b shows a map of the $v_F$
surface for a variety of different Fermi levels, while Fig. 7c (solid line)
shows the minimum value of $\vec q_0\cdot\vec v_F$, as a function of Fermi
energy.  What is plotted is the velocity normalized to a zone boundary phonon,
$v_i^*=\hbar\pi v_{Fi}/a$; for a phonon away from the zone boundary, the curves
should be scaled by the ratio $qa/\pi$.  (In this plot, the renormalized band
dispersion parameters are used, $t=0.25eV$, $\Delta =1eV$.)
{}From these figures, it is clear that
there is only a narrow range of energies for which $\vec q\cdot\vec v_F\le
\omega_{ph}$, particularly near the zone boundary.  Moreover, the slowest
velocities are associated with hopping in directions near X and Y -- i.e.,
towards the Cu's.  For these directions, and sufficiently close to the vHs,
the electronic motion is slow enough that the molecular distortions can keep
up.
\par
The condition
$$\omega_{ph}\simeq\vec q\cdot\vec v_F\eqno(28a)$$
can be reinterpreted in a number of ways.  First, for acoustic phonons the
sound
velocity is $c_s=\omega_{ph}/q$, and Eq. 28a becomes
$$c_s\simeq v_F\eqno(28b)$$
-- the `surf riding' condition that signals the onset of Landau damping.
Finally, for large $q$, the electronic dispersion should be treated more
carefully, and Eq. 28a should be replaced with
$$\hbar\omega_{ph}\simeq E_{\vec k+\vec q/2}-E_{\vec k-\vec q/2},\eqno(28c)$$
for some electronic wave number $k$, where $E_{\vec k}$ is the electronic
energy dispersion.  Equation 28c should be compared to the resonant
denominators
of Eqs. 16a and 21. The right-hand side of Eq. 28c is plotted as the dashed
line
in Fig. 7c, and it can be seen that use of this condition broadens the range of
energies over which the inequality is satisfied.  This last equation is the
condition for a {\it resonant electron-phonon interaction} -- i.e., both energy
and momentum are conserved. Such a coupling will lead to large modifications of
both the electronic and phononic dispersions.
\par
This suggests a new interpretation for the vHs.  Away from a vHs, electronic
hopping is very fast, and it is hard to say just which atom the electron is on
instantaneously.  The band is a mixture of Cu and O, with both at an
intermediate valence -- i.e., the coupling is covalent.  The Born-Oppenheimer
(adiabatic) approximation is valid: the electronic motion is so fast that it
responds instantaneously to any nuclear motions.  Near a vHs, the opposite,
anti-adiabatic limit is valid.  The electronic motions are {\it slow compared
to
the nuclear motions} -- for at least some of the electrons, $v_F=0$.  In this
limit, the {\it valence} of an individual atom is much more definite.  For
instance, in the fast hopping regime, the local distortions associated with
O$^-$ vs O$^{2-}$ do not have a chance to develop, so all O's have a common,
intermediate value of radius.  Near the vHs, on the other hand, Cu-O hopping is
slowed down, allowing local distortions to develop, so that an O$^-$ can be
clearly distinguished from an O$^{2-}$ by its smaller radius.  In short, near
a vHs the nuclear motion can `instantaneously' follow the electronic motion,
and
the bonding goes from covalent to substantially more ionic.  This leads to the
strong enhancement of the electron-phonon coupling.
\par
Since correlation effects also act to slow down the Cu-O hopping, this may be
an explanation for why correlation effects act to pin the vHs nearer to the
Fermi level at half filling.

\subsection{Future Work}

The present paper is the second of a projected series of papers whose aim is to
provide a firm basis for analyzing the vHs-JT model within the three-band
model of the cuprates.  The first paper (XA)\cite{RMX} had two purposes: the
main purpose was to extend the slave boson calculation of correlation effects
to the full 3-band model; the second, to explore a possible linear
electron-phonon coupling induced by spin-orbit coupling.  The present paper
(XB)
further explores electron-phonon coupling in the {\it Peierls} regime, when
correlation effects are expected to be weak.  A strong, linear component is
found to contribute to the dynamic JT fluctuations, leading to important
polaronic effects.
\par
A related paper (XIA)\cite{RMXC}, presently under preparation,
analyzes electron-phonon coupling in the {\it spin-Peierls} regime,
where correlation effects are strong -- in the 2D Heisenberg limit near half
filling.  In future papers, I hope to extend these results to the three-band
model.  At this point, I should be able to study both Peierls and spin-Peierls
regimes in parallel.  Only then will I be able to give a detailed calculation
of
the doping dependence of the structural transitions, including the `pseudogap',
and to better understand the possible nanoscale phase separation in these
materials.
\par
A brief mention should be made about spin-orbit coupling, discussed in XA.
Spin-orbit coupling leads to interesting modifications of Fermi surfaces, and,
since it leads to a linear electron-phonon coupling, it could play a role in
driving structural instabilities. This effect competes with the dynamic JT
effect (a competition well known in molecular JT problems).  It has been
suggested that spin-orbit coupling is too small to drive any structural
instabilities.  However, a simple estimate (Appendix E) shows that, near a vHs,
this suggestion may not be correct.

\section{Conclusions}

The present calculations clarify the nature of electron-phonon coupling in the
HTT-LTO-LTT phase transitions in LSCO, LBCO.  First, the dominant effect in the
transition is the change of Cu-O bondlength with doping.  This large valence
dependence of the oxygen ion radius is a signature of a strong, linear
electron-phonon coupling.  However, as the corresponding phonon modes soften,
they interact with the very soft tilt modes, driving them unstable.  Thus,
although the dominant coupling is linear, the final instability involves
quadratically coupled phonons.
\par
Therefore, the full theory involves both holes and two
types of phonons -- the stretch modes and the tilt modes.  However, the main
effects on the holes are produced by the stretch modes, and including only
these leads to a major simplification within the vHs-JT model,
in that the earlier, quadratic phonon coupling to electrons is
replaced by a linear coupling.  This linear, in-plane coupling can now be
treated by conventional techniques, providing a possible microscopic basis for
the many {\it polaronic models} of superconductivity\cite{pol}.  A striking
prediction of the model is a significant band narrowing, {\it only in the
immediate vicinity of the vHs}, thereby providing a plausible explanation for
the existence of `extended vHs'.
\par
The reason the band narrowing is localized near the vHs is connected with a
special property of the vHs. Near a vHs, the material is in the `ionic limit',
in that the electronic motion is slow compared to phonon motion, so valence
fluctuations are well defined.

\section{Acknowledgements}

I wish to thank T. Egami and I. Bozovic for stimulating conversations, J.B.
Goodenough and M. Braden for providing me with preprints of their work, and
the organizers of the U. Miami Workshop on High T$_c$ Superconductivity for
providing a most stimulating atmosphere.  Publication 642 of the Barnett
Institute.

\appendix

\section{Doping Dependence of LTO Transition}

Figure 3 of Ref. \cite{RM8B} showed that the phonon softening associated with
the LTO transition in La$_2$CuO$_4$ could be modelled in terms of a microscopic
picture involving phonon anharmonicity and electron-phonon coupling.  The data
were insufficient to fix values for all the parameters, and in particular
the electron-phonon interaction parameter could be varied over a large range,
including zero.  The same model can be applied to explaining the doping
dependence of the transition temperature, $T_{LTO}$, for either Sr or Nd
doping, Figure 2.
\par
{}From Eq. 35 of Ref. \cite{RM8B}, $T_{LTO}$ is the solution of
$$\omega_0^2+[9\Gamma_1+\Gamma_2-\alpha_e^2N(0)]\Delta_1=0.\eqno(A1)$$
Here, $\omega_0$ is the `bare' phonon frequency, the $\Gamma_i$'s represent
anharmonic (quartic) phonon-phonon couplings, $\alpha_e$ is an electron-phonon
coupling parameter, and $\Delta_1\propto coth(\omega_1/2k_BT)$ is proportional
to the fluctuations of the tilt amplitude (the frequency $\omega_1$ is defined
in Ref. \cite{RM8B}).  In the absence of electron-phonon coupling, Eq. A1 has a
solution only if $\omega_0^2$ is negative.  Such negative values can be caused
by strains due to interlayer mismatch, and to explain the doping dependence of
$T_{LTO}$, it will be assumed that Eq. 2 holds, which can be rewritten as
$$\omega_0^2=\tilde\omega_0^2-\tilde\Gamma (a-a_2).\eqno(A2)$$
(Since $n_h$ only varies from 1 to $\sim 1.2$, the main effect of the middle
term in Eq. 2 can be absorbed into a renormalized value of $\tilde\omega_0$.)
\par
The temperature dependence of Eq. A1 comes from that of $\Delta_1$ and that of
$N(0)$, whereas the doping ($x$) dependence is assumed to come from that of
the lattice mismatch, $(a-a_2)$, and that of $N(0)$.  The lattice mismatch is
estimated in Appendix B.  The hardest factor to
quantify is the dos, $N(0)$, due to the important role of correlation effects.
In the absence of correlation effects, assuming a rigid-band doping dependence
and a vHs shifted away from half filling, $N(0)$ can be written as
$$N(0)={1\over 2B}ln\bigl({B\over\sqrt{\mu^2+(k_BT)^2}}\bigr),\eqno(A3)$$
where $B$ is the bandwidth, and $\mu (x)$ is the distance in energy between the
vHs and the Fermi level.  This equation should hold in the doped materials near
optimum $T_c$, but near half filling correlation effects become important.
\par
Corelation effects reduce the electron-phonon coupling, $\alpha_e$, for the
usual Peierls interaction\cite{KiT,RMX}, but there is also a spin-Peierls
coupling, for electron-phonon coupling stronger than a critical
threshold\cite{TH,RMXC}.  The strength of this effect is comparable at half
filling and at the vHs: while nesting is better at half filling (the {\it
pseudofermions} have a square Fermi surface), vHs nesting is suppressed by
matrix element effects\cite{RMXC}.  Hence, the conventional nesting at half
filling will be of comparable strength to the vHs nesting in the doped
material.
Therefore, for the calculations of $T_{LTO}$ in Fig. 2, a simplifying
assumption
is made -- that the vHs is pinned to the Fermi level, but that the strength of
coupling (represelted by $N(0)$) is independent of doping.
\par
In analyzing the $T$-dependence of the soft mode frequency in the undoped
La$_2$CuO$_4$\cite{RM8B}, it was found that the transition could be equally
well
fit by either a purely anharmonic model or one with a strong component of
electron-phonon coupling.  In analyzing the doping dependence, it is found that
the same ambiguity of parameter values remains, and Fig. 2a-c correspond to
essentially the same parameters as Fig. 3a-c of Ref. \cite{RM8B}.
Identical parameters cannot be used, since the transition temperature at
zero doping is considerably higher for the present data than for the earlier
(the fits in Ref. \cite{RM8B} found $T_{LTO}$ between 455-508K).  There is no
unique technique for adjusting the parameter values to the new $T_{LTO}$
without
fitting the entire transition, for which data are not available.  However, $T_
{LTO}$ is directly related to the `bare' phonon frequency, $\omega_0$, so this
parameter was increased slightly (by $\sim 4-8\%$) to correct for the higher $T
_{LTO}$ values (corresponding to fit values of $T_{LTO}\simeq 540-576K$ at
$x=0$).
\par
The parameters i$\omega_0$, $\tilde\omega_0$, and $\tilde\Gamma$ are listed in
Table I.  The other parameters are the same as in the respective parameter sets
of Table II of Ref. \cite{RM8B}.  Note that the lattice mismatch is clearly
responsible for the instability: whereas $\omega_0^2$ is negative, signalling
an
instability, the bare parameter $\tilde\omega_0$ is large and positive.
\par
One feature should be noted: the renormalized Nd parameter, $(a-a_2)^*$ from
Appendix B, had to be used, to get good fits.  This is directly related to
the fact that the electron-phonon coupling term was assumed to be doping
independent.  If $\alpha_e^2N(0)$ is assumed to be smaller at $x=0$, then
electron-phonon coupling would favor a higher $T_{LTO}$ in the doped material,
a trend that would have to be compensated by a larger value of $\tilde\Gamma$.
To explain the Nd-doping results then would require a smaller value of
$(a-a_2)$
when $y$ is varied.  Contrawise, if $\alpha_e^2N(0)$ increases as half filling
is approached, a larger value of the y-coefficient of $(a-a_2)$ is needed.
This result can be turned around: if we knew perfectly the theoretical form of
$(a-a_2)$, we could determine from experiment the precise doping dependence of
the electron-phonon coupling coefficient.  Unfortunately, the uncertainty in
parameters is too large to determine whether $\alpha_e^2N(0)$ increases or
decreases with doping.

\section{Atomic Radii}

In testing an ionic model for the LTO transition, it is necessary to estimate
the effective ionic radii in the various layers as a function of doping.
This will be done in two ways: by using tabulated values of radii\cite{Shan}
and
via valence bond sums\cite{BrAlt}.  Exact agreement with experiment should
not be expected, since the tabulated values are averages over many compounds.
Nevertheless, an approximate agreement is found.

\subsection{On the LaO Plane}

Commensurability strains will arise in LSCO whenever the tolerance
factor, Eq. 1, is different from unity.  It should be noted that, in this
equation, both $d_{LaO}$ and $d_{CuO_2}$ are functions of doping.  In undoped
La$_2$CuO$_4$, the CuO$_2$ planes are under compression, $t<1$.  The atoms in
the LaO layer are easier to discuss, since they can be treated in the ionic
limit.  The radii will first be estimated from Shannon's tables\cite{Shan},
assuming the La site to be 9-fold coordinated\cite{Haz}.  In this case,
assuming the conventional O$^{2-}$ radius, $r_{O^{2-}}=1.40\AA$, the relevant
radii are $r_{La^{3+}}=1.216\AA$, $r_{Nd^{3+}}=1.163\AA$,
$r_{Ba^{2+}}=1.47\AA$,
$r_{Sr^{2+}}=1.31\AA$, and $r_{Ca^{2+}}=1.18\AA$.  Thus, for undoped
La$_2$CuO$_
4$, $d_{LaO}=2.616\AA$, while in an Sr and Nd co-doped material, La$_{2-x-y}
$Sr$_x$Nd$_y$CuO$_4$, $d_{LaO}=(2.616+0.047x-0.027y)\AA$.
Thus, Sr doping reduces the interlayer strains, while Nd substitution makes
them worse.
\par
In the $T^{\prime}$ structure, the rearrangement of the apical O's to
interstitial positions expands the lattice, and leaves the CuO$_2$ layers under
tension\cite{BGood}, and leading to materials which can be electron doped.
These structures will not be further discussed here.

\subsection{On the CuO$_2$ Plane}

Shannon's tables list $r_{Cu^{1+}}=0.77\AA$, $r_{Cu^{2+}}=0.73\AA$, and
$r_{Cu^{3+}}=0.54\AA$.  Assuming the Cu is all in the 2+ state, this yields
$d_{CuO_2}=2.13\AA$, to be compared with $a_2\equiv d_{LaO}/\sqrt{2}=1.850\AA$
in La$_2$CuO$_4$.  This severe mismatch is largely corrected for by a large
{\it
molecular} LT effect in the CuO$_6$ octahedra: the electronic degeneracy of the
Cu d levels ($d_{x^2-y^2}$, $d_{z^2}$) is split by a molecular distortion in
which the planar CuO distance becomes much smaller than the apical distance.
A rough estimate of the effect can be made from the Shannon values.  For Cu$^{
2+}$, the one hole in the Cu d-shell is generally an equal mix of the $d_{x^2-
y^2}$ and the $d_{z^2}$ orbitals.  In the JT state, the hole is completely
taken from the $d_{x^2-y^2}$ orbital.  Hence, for in-plane lengths, the
relevant
orbital has the same occupation as in the isotropic Cu$^{3+}$ ion, with a
corresponding length $d_{CuO_2}^{in-plane}\simeq 1.94\AA$.  For the apical
lengths, the $d_{z^2}$ orbital has the same length as in Cu$^{1+}$, or
$d_{CuO_2}^{apical}\simeq 2.17\AA$.  Experimentally, these lengths are found to
be $d_{CuO_2}^{in-plane}=1.905\AA$ and $d_{CuO_2}^{apical}=2.421\AA$ in
La$_2$CuO$_4$ at 10K.  For the apical O, the agreement could be further
improved
by noting that screening is very weak along the c-axis, so the Pauling
value\cite{Pau,Shan} $r_{Cu^{1+}}=0.96\AA$ is more appropriate, yielding $d_{C
uO_2}^{apical}\simeq 2.36\AA$.
\par
To extend the above picture to the doped regime, it is necessary to understand
where the doped holes go.  Experimentally, it is found that hole occupation of
the Cu $d_{z^2}$ orbital only becomes apparent for $x>0.15$ -- i.e., just above
the optimal doping for superconductivity.  Hence, in the low-doping regime it
can be assumed that the holes go primarily onto the in-plane O's.  Chemically,
this would correspond to the appearance of the species $O^{1-}$.  This is a
rare
species, and its ionic radius is not well established.  The present analysis
will provide an estimate of its value.

\subsection{Doping Dependence of T$_{LTO}$}

Using the above estimates, in La$_2$CuO$_4$, $a_2=1.850\AA$, $a_1\equiv
d_{CuO_2
}=1.94\AA$, and $a=1.905\AA$, for a reasonably consistent picture.  The doping
dependence of $a_2$ is
$$a_2=(1.850+0.033x-0.019y)\AA,\eqno(B1)$$
while, experimentally\cite{Breu}
$$a=(1.901-0.085x+0.0y)\AA\eqno(B2)$$
so
$$a-a_2=(0.051-0.118x+0.019y)\AA\eqno(B3)$$
at room temperature.  Note that there are additional small corrections, which
are neglected.  In particular, no correction is made for the fact that the La
and the O are not strictly coplanar (and similarly for the Cu and O).  Also,
there is no temperature dependence of the lattice constants, and all 9 La-O
distances are assumed to be the same.
\par
The above model predicts too small a change of $a-a_2$ with $y$ to agree with
experiment.  Empirically, the ratio of the coefficients of $y$ and $x$ can be
estimated from the change in $T_c$ with these parameters.  Now
$$T_{LTO}=T_0-T_xx+T_yy,\eqno(B4)$$
with $T_0\simeq 546K$, $T_x\simeq 2456K$ (from Ref. \cite{Tin}), and $T_y\simeq
560K$ (Ref. \cite{Buch}).  This yields $T_x/T_y=4.39$, compared to the
predicted
value, $0.118/0.019=6.21$.  In the following subsection, we will see that this
lattice mismatch can also be estimated from valence bond sums, and here the
$y$-coefficient is found to be too large.  In the present paper, an effective
average value will be assumed, to give better agreement with experiment:
$$(a-a_2)^*=(0.051-0.118x+0.027y)\AA.\eqno(B5)$$
\par
Note that the calculated transition (Fig. 2) terminates at a doping $x\simeq
0.21$, where $\omega_0^2$ becomes positive (Eq. 2), while the interlayer strain
vanishes only when $a=a_2$, at a much larger doping, $x=0.43$ when $y=0$.

\subsection{Valence Bond Sums}

In view of the importance of the above results, it is preferable to recalculate
$a_2$ by another technique.  The method of valence bond sums\cite{BrAlt} has
been applied by several authors to the high-T$_c$
superconductors\cite{deL,Ta,Tan}.  In this model, the valence associated with
a bond between atoms A and B is written as
$$V_{AB}=exp\bigl({r_{AB0}-r_{AB})\over r_0}\bigr),\eqno(B6)$$
where $r_0=0.37\AA$ is a constant, $r_{AB0}$ is tabulated for over a hundred
possible pairs\cite{BrAlt}, and $r_{AB}$ is the measured AB distance.  The
total
valence of atom A is then
$$V_A=\sum_iV_{AB_i},\eqno(B7)$$
where the sum is over all close neighbor atoms, $B_i$.  This can be rearranged
to give the ideal radius for a given atom:
$$r_{AB}=r_{AB0}+r_0ln({N_A\over V_A}),\eqno(B8)$$
where $N_A$ is the coordination of atom A ($N_{La}=9$).  Taking $V_{La}=V_{Nd}=
3$ and $V_{Sr}=2$, and using tabulated values\cite{BrAlt} for $r_{AB0}$, it
is found that
$$a_2|_{val.bond}=(1.823+0.069x-0.047y)\AA,\eqno(B9)$$
so
$$(a-a_2)_{val.bond}=1.31(0.060-0.118x+0.036y)\AA.\eqno(B10)$$
The coefficient of $y$ is here found to be much larger than that estimated
from the Shannon tables, and even larger than the experimental value.

\subsection{The Radius of O$^-$}

Now, from Eq. 3b,
$$(a_1-a)={2C_2\over C_1}(a-a_2).\eqno(B11)$$
Since $a_1$ should be independent of $y$, the experimental observation that
$a$ is approximately independent of $y$ requires that $C_2<<C_1$.  For
example, if $2C_2/C_1\simeq 0.1$, then $a$ would have to include a term $\sim
0.003y$.  This would then yield
$$a_1\simeq (1.904-0.097x)\AA.\eqno(B12)$$
(Note that, at $x=0$, this yields a smaller value of $a_1$ than the chemical
estimate, $a_1\simeq 1.94\AA$, made above.)
Combining this with Eq. B1 yields a tolerance factor
$$t={a_2\over a_1}\simeq 0.972+0.069x-0.010y,$$
or $t=1$ at $x=0.42$, when $y=0$.

If all the doping is assumed to go into the oxygens, then setting $x=2$ is
equivalent to converting both planar O's to O$^-$, yielding
$r_{O^-}=(1.40-0.194)=1.206\AA$.  Grenier\cite{Gren} has postulated that
O$^{2-}$ is too large to readily diffuse, and that the diffusing species in
YBCO and the other cuprates is probably O$^-$.  He estimates its radius to
be $\sim 1.1\AA$, comparable to the present estimate.  From the valence bond
sums, a similar estimate can be made, $r_{O^{2-}}-r_{O^-}\simeq r_0ln2 \simeq
0.256\AA$.
\par
The large increase in radius on going to $O^{2-}$ is consistent with the
suggestion that $O^{2-}$ is inherently unstable, being stabilized in a lattice
by the Madelung potential\cite{Bilz}.  This near instability leads to a large,
nonlinear polarizability which has been hypothesized to lead to
ferroelectricity
in the perovskites\cite{Bilz}.

\section{In-Plane Mode Coupling in a Tight Binding Model}

In this Appendix, the electron-phonon coupling associated with the various
structural distortions will be estimated in a tight-binding model, assuming
that the coupling arises from the bond-length dependence of the hopping
parameter, $t_{CuO}$.  In the standard three-band model, assuming $t_{OO}=0$
for simplicity, a very general case can be handled.  Assume that the
distortions have led to a $\sqrt{2}\times\sqrt{2}$ unit cell, with alternating
hopping parameters $t_1$, $t_2$, $t_3$, and $t_4$ along the $x$-axis (Fig.
3a,c), and $t$ along $y$. Then four possible distortion modes are (1) Cu-Cu
bond stretching mode, $t_1=t_2=t-\delta t$, $t_3=t_4=t+\delta t$; (2) O-O bond
stretching mode, $t_1=t_4=t-\delta t$, $t_2=t_3=t+\delta t$; (3) ferroelectric
(FE) mode, $t_1=t_3=t-\delta t$, $t_2=t_4=t+\delta t$; (4) shear strain, $t_1=
t_2=t_3=t_4=t-\delta t$. For the shear strain mode, the y-hopping must be reset
to $t_y=t+\delta t$, and for comparable distortions, $\delta t$ must be smaller
by a factor of $\sqrt{2}$ than for models 1-3.
\par
For these distortions, the energy dispersion is given by the solutions of
$$(W-W_{\vec k})(W-W_{\vec k-\vec q})=\Gamma ,\eqno(C1)$$
with $W=E(E-\Delta )$,
$$W_{\vec k}=4t^2s_y^2+{\sum_{i=1,4}t_i^2\over 2}-(t_1t_2+t_3t_4)\bar c_x,
\eqno(C2)$$
$$\Gamma ={1\over 4}[(t_1^2-t_3^2)^2+(t_2^2-t_4^2)^2+2(t_1t_4-t_2t_3)^2$$$$
+2(t_1t_2-t_3t_4)^2(1-2\bar c_x^2)],\eqno(C3)$$
and $s_i=sin(k_ia/2)$, $c_i=cos(k_ia/2)$, $\bar c_x=cos(k_xa)$.
The values of $W_{\vec k-\vec q}$ are also given by Eq. C2, with the
substitution $\vec k\rightarrow\vec k-\vec q$ in the arguments of the trig
functions.  In particular, for $\vec q=\vec q_0$ (Fig. 3c-f) $s_i\rightarrow -
c_i$, $c_i\rightarrow s_i$, $i=x,y$.  The same dispersion relations also hold
for the distortions of Fig. 3a,b, but with $\vec q=\vec q_1$.  In this case,
the terms in $s_y$ remain unchanged in $W_{\vec k-\vec q}$.  The calculated
energy dispersions are illustrated in Fig. 5, assuming $t=1eV$, $\Delta =4eV$,
and $\delta t=0.1eV$. Note that the Cu-Cu and O-O bond stretching modes give
identical dispersions for a $\vec q_1$ distortion (Fig. 5a,b -- this can be
shown analytically from Eq. C1) but very different dispersions for the $\vec
q_0$ distortion (Fig. 5c,d).
\par
In Eq. C3, $\Gamma$ is the square of an umklapp scattering gap.  For the
four various modes, it can be simplified to
$$\Gamma_i=\cases{64t^2\delta t^2s_x^2c_x^2& if i=1;\cr
                  16t^2\delta t^2& if i=2;\cr
                  0& if i=3,4.\cr}\eqno(C4)$$
Note that there is no gap for the FE mode, but a large gap $\sim \delta t$
opens up for the other modes, although it vanishes at the vHs for the Cu-Cu
bond
stretching mode.  When $\gamma =0$, the dispersion becomes $E=E_{\vec k}$, with
$$E_{\vec k}={\Delta\over 2}+\sqrt{({\Delta\over 2})^2+W_{\vec k}}.\eqno(C5)$$
For $\Gamma\ne 0$, by defining $F(E)=E-\Delta /2$, Eq. C1 can be rewritten
$$F^2(E)={F^2(E_{\vec k})+F^2(E_{\vec k-\vec q})\over 2}
$$$$\pm\sqrt{({F^2(E_{\vec
k})-F^2(E_{\vec k-\vec q})\over 2})^2+\Gamma}.\eqno(C6)$$
\par
Explicitly, for either the O-O bond stretching mode or the FE mode,
$$W_{\vec k}=4t^2(s_x^2+s_y^2)+4\delta t^2c_x^2,\eqno(C7a)$$
for the Cu-Cu bond stretching mode:
$$W_{\vec k}=4(t^2+\delta t^2)s_x^2+4t^2s_y^2,\eqno(C7b)$$
and for the strain:
$$W_{\vec k}=4(t-\delta t)^2s_x^2+4(t+\delta t)^2s_y^2.\eqno(C7c)$$
\par
When $E_{\vec k}-E_{\vec k-\vec q}$ is large, the dispersion is essentially
given by Eq. C5.  However, when $E_{\vec k}=E_{\vec k-\vec q}$, the
excitation spectrum is gapped.  For the O-O bond stretching mode,
$$E\simeq E_{\vec k}\pm{2t\delta t\over\sqrt{({\Delta\over 2})^2+4t^2(s_x^2+s_y
^2)}},\eqno(C8)$$
to lowest order in $\delta t$.  Since $s_x^2+s_y^2=1$ everywhere on the Fermi
surface, the gap is isotropic, as seen in Fig. 5d.  Similar equations hold for
the remaining two distortions, except that the factor $2t$ in the numerator
becomes $8ts_xc_x$ for case 1 (Cu-Cu hopping) and $2\sqrt{2}t(s_x^2-s_y^2)$ for
case 4 (shear strain).
Even though there is no gap in the FE band, the distortion still leads to a net
energy lowering, but it is $\sim\delta t^2$.  This is a 2D version of the model
discussed by Egami, et al.\cite{Eg}.
\par
The gap at the vHs can be calculated as
$$\Delta E_i={2\beta_i\delta t\over\sqrt{1+({\Delta\over 4t})^2}},\eqno(C9)$$
$i=1,4$ labelling the four modes, with $\beta_1=\beta_3=0$, $\beta_2=1$, and
$\beta_4=\sqrt{2}$.  For the strain mode, this represents a splitting of the
vHs degeneracy at the $X$ and $Y$ points.

\section{Polaronic Calculations}

\subsection{Canonical Transformation: Full Three Band Model}

The complete, fully 2D solution to Eqs. 11-14 involves a system of 18 equations
in 18 unknowns, the $f$'s and $h$'s.  Ten unknowns
can immediately be eliminated by direct substitution.  Defining $\tilde
f_i=f_i/
\omega_{\vec q}$, these are
$$h_9={t_{x\vec k-\vec q}\tilde f_4+t_{y\vec k}\tilde f_1\over\hbar\omega_{\vec
q}},\eqno(D1a)$$
$$h_8={t_{x\vec k}\tilde f_3+t_{y\vec k-\vec q}\tilde f_2\over\hbar\omega_{\vec
q}},\eqno(D1b)$$
$$h_7={t_{y\vec k}\tilde f_3+t_{y\vec k-\vec q}\tilde f_4\over\hbar\omega_{\vec
q}},\eqno(D1c)$$
$$h_6={t_{x\vec k}\tilde f_1+t_{x\vec k-\vec q}\tilde f_2\over\hbar\omega_{\vec
q}},\eqno(D1d)$$
$$h_5=-{[t_{x\vec k-\vec q}\tilde f_1+t_{x\vec k}\tilde f_2
+t_{y\vec k-\vec q}\tilde f_3+t_{y\vec k}\tilde f_4]\over\hbar\omega_{\vec
q}},\eqno(D1e)$$
with the five equations formed by interchanging $h_i\leftrightarrow\tilde f_i$.
\par
Solution of the remaining eight equations is more involved.  Defining column
matrices
$$\tilde H=\left(\matrix{h_1\cr
                         h_2\cr
                         h_3\cr
                         h_4\cr}\right),\eqno(D2a)$$
$$\tilde F=\left(\matrix{\tilde f_1\cr
                         \tilde f_2\cr
                         \tilde f_3\cr
                         \tilde f_4\cr}\right),\eqno(D2b)$$
$$\tilde G=\left(\matrix{\hbar^2\tilde g_{\vec q}(\vec k)\cr
                         0\cr
                         0\cr
                         0\cr}\right),\eqno(D2c)$$
then the remaining equations become
$$\tilde A\tilde H+\tilde B\tilde F=\tilde G,\eqno(D3a)$$
$$\tilde A\tilde F+\tilde B\tilde H=0,\eqno(D3b)$$
with solution
$$\tilde H={1\over 2}[(\tilde A+\tilde B)^{-1}+(\tilde A-\tilde B)^{-1}]\tilde
G,\eqno(D4a)$$
$$\tilde F={1\over 2}[(\tilde A+\tilde B)^{-1}-(\tilde A-\tilde B)^{-1}]\tilde
G.\eqno(D4b)$$
Introducing the notations $a=t_{x\vec k-\vec q}$, $b=t_{x\vec k}$, $c=t_{y\vec
k-\vec q}$, $d=t_{y\vec k}$, and $e=\hbar\omega_{\vec q}\Delta$, the matrices
$\tilde A=\tilde P\tilde R\tilde P$ and $\tilde B$ can be written
$$\tilde B=\left(\matrix{e&0&0&0\cr
                         0&-e&0&0\cr
                         0&0&e&0\cr
                         0&0&0&-e\cr}\right),\eqno(D5a)$$
$$\tilde P=\left(\matrix{a&0&0&0\cr
                         0&b&0&0\cr
                         0&0&c&0\cr
                         0&0&0&d\cr}\right),\eqno(D5b)$$
$$\tilde R=\left(\matrix{\alpha&2&1&2\cr
                         2&\beta&2&1\cr
                         1&2&\gamma&2\cr
                         2&1&2&\delta\cr}\right),\eqno(D5c)$$
with
$$\alpha ={W-d^2\over b^2},\eqno(D6a)$$
$$\beta ={W-c^2\over a^2},\eqno(D6b)$$
$$\gamma ={W-b^2\over d^2},\eqno(D6c)$$
$$\delta ={W-a^2\over c^2},\eqno(D6d)$$
and $W=\hbar^2\omega_{\vec q}^2+a^2+b^2+c^2+d^2$.
\par
Defining $\tilde H_{\pm}=(\tilde A\pm\tilde B)^{-1}\tilde G$, $W_{\pm}=W\pm e$,
${\cal A}=a^2+c^2$, ${\cal B}=b^2+d^2$, the explicit solution is
$$h_{i\pm}={\hbar^2\tilde g_{\vec q}(\vec k)\hat h_{i\pm}\over
(W_+W_--4{\cal A}{\cal B})(W_{\pm}-{\cal A})},\eqno(D7a)$$
$$\hat h_{1\pm}=W_{\mp}(W_{\pm}-{\cal A})+c^2(W_{\mp}-4{\cal B}),\eqno(D7b)$$
$$\hat h_{2\pm}=-2ab(W_{\pm}-{\cal A}),\eqno(D7c)$$
$$\hat h_{3\pm}=-ac(W_{\mp}-4{\cal B}),\eqno(D7d)$$
$$\hat h_{4\pm}=-2ad(W_{\pm}-{\cal A}).\eqno(D7e)$$
The most important factor in the above expressions is the first factor in the
denominator of Eq. D7a, which can be rewritten
$$W_+W_--4{\cal A}{\cal B}=\hbar^2\omega_{\vec q}^2[\hbar^2\omega_{\vec q}^2
+2({\cal A}+{\cal B})-\Delta^2]+({\cal A}-{\cal B})^2$$
$$=[\hbar^2\omega_{\vec q}^2-(E_{\vec k}-E_{\vec k-\vec q})^2]
[\hbar^2\omega_{\vec q}^2-(E_{\vec k}+E_{\vec k-\vec q}-\Delta)^2].\eqno(D7f)$$
Here, the first term in brackets contains the essential $\omega$-dependence of
the problem, and $\omega_{\vec q}$ can essentially be set to zero in all
remaining terms.
\par
The residual interaction terms are given by Eq. 17, with
$$H_{e-e}=-\sum_{\vec k,\vec k^{\prime},\vec q}\tilde
g_{\vec q}(\vec k)\rho_{1+,\vec k,\vec q}B_{\vec k^{\prime},\vec q},\eqno(D8)$$
and
$$H_{e-ph}^{(2)}=-\sum_{\vec k,\vec q,\vec q^{\prime}}{Q_{\vec q}Q_{\vec q^{
\prime}}\tilde g_{\vec q}(\vec k)\over\hbar}\times$$$$
\times\bigl(
f_{1\vec k,\vec q^{\prime}}\rho_{6-,\vec k-\vec q,\vec q^{\prime}-\vec q}
-f_{2\vec k+\vec q^{\prime},\vec q^{\prime}}\rho_{6-,\vec k-\vec q,
-\vec q^{\prime}-\vec q}$$$$
+f_{3\vec k,\vec q^{\prime}}\rho_{8-,\vec k-\vec q,\vec q^{\prime}-\vec q}
-f_{4\vec k+\vec q^{\prime},\vec q^{\prime}}\rho_{8-,\vec k-\vec q,
-\vec q^{\prime}-\vec q}$$$$
+f_{5\vec k,\vec q^{\prime}}\rho_{2-,\vec k-\vec q,\vec q^{\prime}-\vec q}
+f_{5\vec k+\vec q^{\prime},\vec q^{\prime}}\rho_{2-,\vec k-\vec q,
-\vec q^{\prime}-\vec q}$$$$
+f_{2\vec k-\vec q,\vec q^{\prime}}\rho_{5-,\vec k,\vec q^{\prime}+\vec q}
-f_{1\vec k+\vec q^{\prime}-\vec q,\vec q^{\prime}}\rho_{5-,\vec k,
\vec q-\vec q^{\prime}}$$$$
+f_{6\vec k-\vec q,\vec q^{\prime}}\rho_{1-,\vec k,\vec q^{\prime}+\vec q}
+f_{6\vec k+\vec q^{\prime}-\vec q,\vec q^{\prime}}\rho_{1-,\vec k,
\vec q-\vec q^{\prime}}$$$$
+f_{8\vec k-\vec q,\vec q^{\prime}}\rho_{3-,\vec k,\vec q^{\prime}+\vec q}
+f_{9\vec k+\vec q^{\prime}-\vec q,\vec q^{\prime}}\rho_{3-,\vec k,
\vec q-\vec q^{\prime}}
\bigr),\eqno(D9)$$
where it is assumed that the terms linear in $Q_qP_q$ will average to zero.

\subsection{Canonical Transformation: Quasi-1D Approximation}

\par
Due to the complicated nature of the solution, Eqs. D1,7, this subsection
introduces a simpler, {\it quasi-1D} approximation: the commutator of $S$ with
the hopping Hamiltonian along the y-direction will be neglected.  The next
subsection gives a different, effective mass type approximation.
\par
In the quasi-1D case, only the $f_i$'s and $h_i$'s with $i=1,2,5,6$ are
non-vanishing.  When $c=d=0$, Equations D7 reduce to $\hat h_{3\pm}=\hat h_{4
\pm}=0$, $f_2=0$,
$$\tilde f_1={-eh_1\over W},\eqno(D10a)$$
$$h_2={-2abh_1\over W},\eqno(D10b)$$
and
$$h_1={\hbar^2\tilde g_{\vec q}(\vec k)W\over W_+W_--4{\cal
A}{\cal B}}.\eqno(D10c)$$
Equation D10a can be rewritten in the suggestive form
$$f_{1\vec k,\vec q}={\hbar\tilde g_{\vec q}(\vec k)\beta_{\vec k,\vec q}\over
\hbar^2\omega_{\vec q}^2-(E_{\vec k}-E_{\vec k-\vec q})^2},\eqno(D11a)$$
with
$$\beta_{\vec k,\vec q}={\hbar^2\omega_{\vec q}^2\Delta\over
(E_{\vec k}+E_{\vec k-\vec q}-\Delta)^2-\hbar^2\omega_{\vec q}^2},\eqno(D11b)$$
and
$$E_{\vec k}={\Delta\over 2}+\sqrt{{\Delta^2\over 4}-t^2_{x\vec k}}.
\eqno(D11c)$$
\par
The effective electron-electron interaction $H_{ee}$ can still be written in
the
form of Eq. D8, while the residual electron-phonon interaction becomes
$$H_{e-ph}^{(2)}=-\sum_{\vec k,\vec q,\vec q^{\prime}}{Q_{\vec q}Q_{\vec q^{
\prime}}\tilde g_{\vec q}(\vec k)\over\hbar}\times$$$$
\times\bigl(
f_{1\vec k,\vec q^{\prime}}\rho_{6-,\vec k-\vec q,\vec q^{\prime}-\vec q}
-f_{1\vec k+\vec q^{\prime}-\vec q,\vec q^{\prime}}\rho_{5-,\vec k,
\vec q-\vec q^{\prime}}$$$$
+f_{5\vec k,\vec q^{\prime}}\rho_{2-,\vec k-\vec q,\vec q^{\prime}-\vec q}
+f_{5\vec k+\vec q^{\prime},\vec q^{\prime}}\rho_{2-,\vec k-\vec q,
-\vec q^{\prime}-\vec q}$$$$
+f_{6\vec k-\vec q,\vec q^{\prime}}\rho_{1-,\vec k,\vec q^{\prime}+\vec q}
+f_{6\vec k+\vec q^{\prime}-\vec q,\vec q^{\prime}}\rho_{1-,\vec k,
\vec q-\vec q^{\prime}}
\bigr).\eqno(D12)$$
Equation D12 can be further simplified, by noting that in the most important
case, $\vec q^{\prime}=-\vec q$ and $2\vec q$ is a reciprocal lattice vector,
and hence equivalent to zero. In this case,
$$H_{e-ph}^{(2)}=-\sum_{\vec k,\vec q}{Q_{\vec q}Q_{-\vec q}
\tilde g_{\vec q}(\vec k)\over\hbar}\times$$$$
\times\bigl(
f_{1\vec k,-\vec q}[\rho_{6-,\vec k-\vec q,0}-\rho_{5-,\vec k,0}]$$$$
+[f_{6\vec k-\vec q,-\vec q}
+f_{6\vec k,-\vec q}]\rho_{1-,\vec k,0}$$$$
-[f_{5\vec k,-\vec q}
+f_{5\vec k-\vec q,-\vec q}]\rho_{1-,\vec k-\vec q,0}
\bigr).\eqno(D13)$$
Note that the term in $\rho_6$ is of the form of Eq. 27, but with a quadratic
$Q$ dependence.
\par
In mean field theory, the four-operator terms are split up according to
$$AB\simeq A<B>+<A>B-<A><B>,\eqno(D14)$$
where $A$, $B$ are products of two operators, and $<\cdot\cdot\cdot >$ denotes
a thermal equilibrium average.  At sufficiently high temperatures, only $Q_{
\vec q}Q_{-\vec q}$ and $\rho_{i\vec k,0}$, $i=5,7$ have non-zero averages.  As
the temperature is lowered, other quantities may develop nonvanishing averages,
signaling a phase transition. A discontinuous change in $\rho_6-\rho_5$
would signal an ionic-to-covalent transition.
\par
This canonically transformed Hamiltonian contains information on many physical
processes, not all of them relevant to the present discussion.  Here, I will
briefly survey some of the terms which arise at mean field level.  From
$H_{e-ph}^{(2)}$, the term in $<Q_{\vec q}Q_{-\vec q}>f_1$ leads to a polaronic
renormalization of the Cu-O$_x$ hopping:
$$t_{x\vec k}\rightarrow t_{x\vec k}(1-\sum_{\vec q}{<Q_{\vec q}Q_{-\vec q}>
\tilde g_{\vec q}(\vec k)h_{1\vec k,-\vec q}\over\hbar^2})\eqno(D15a)$$
(neglecting terms in $\vec k-\vec q$).  The average over $Q^2$ is given by
$$\omega_{\vec q}^2<Q_{\vec q}Q_{-\vec q}>=\hbar\omega_{\vec q}(N_{\vec q}+{1
\over 2}),\eqno(D15b)$$
where $N_{\vec q}$ is given by Eq. 25b.
There is also a renormalization of the phonon frequency,
$$\omega_{\vec q}^{\prime 2}=\omega_{\vec q}^2-\sum_{\vec k}{2f_{1\vec k,-
\vec q}\tilde g_{\vec q}(\vec k)<\rho_{6-,\vec k-\vec q,0}-\rho_{5-,\vec k,0}>
\over\hbar}.\eqno(D16)$$
The presence of $\rho_6-\rho_5$ indicates that this softening is associated
with
charge transfer effects, and will be modified by correlation effects: in the
insulating phase at half filling, $\rho_5$ is constrained to be unity.
\par
In $H_{e-e}$, the mixed terms will lead to additional interference effects.
However, the most important term is given by Eq. 18 in the main text.

\subsection{Canonical Transformation: One Band Approximation}

An alternative approximate solution to Eq. 11 can be found by assuming that
only
the antibonding band is populated by holes, and projecting the electron-phonon
Hamiltonian onto this band.  The electronic Hamiltonian is
$$H_e=\sum_{\vec k}E_{\vec k}\psi^{\dagger}_{\vec k}\psi_{\vec k},\eqno(D17)$$
with $E_{\vec k}$ given by Eq. 6,
$$H_{e-ph}=\sum_{\vec k}Q_q\alpha_{\vec k}\psi^{\dagger}_{\vec k+\vec q}\psi_{
\vec k},\eqno(D18a)$$
and, at the Fermi surface,
$$\alpha_{\vec k}\simeq{i\tilde g(\vec k)t(s_x-s_y)\over\sqrt{({\Delta\over 2})
^2+4t^2(s_x^2+s_y^2)}}.\eqno(D18b)$$
\par
In this case, by the same calculations as for the quasi-1D case, we find
$$S=i\sum_{\vec k}[f_{1\vec k}Q_q-if_{2\vec k}P_q]\psi^{\dagger}_{\vec k+\vec
q}
\psi_{\vec k},\eqno(D19)$$
$$f_{1\vec k}=-{(E_{\vec k+\vec q}-E_{\vec k})f_{2\vec k}\over\hbar},
\eqno(D20a)$$
$$f_{2\vec k}={-\hbar^2\alpha_{\vec k}\over (E_{\vec k+\vec q}-E_{\vec k})^2-
\hbar^2\omega_q^2}.\eqno(D20b)$$
$$H_{e-e}=-\sum_{\vec k,\vec k^{\prime}}\alpha_{\vec k}f_{2\vec k^{\prime}}
\psi^{\dagger}_{\vec k^{\prime}+\vec q}\psi_{\vec k^{\prime}}\psi^{\dagger}_{
\vec k+\vec q}\psi_{\vec k},\eqno(D21)$$
and $H_{e-ph}^{(2)}$ is given by Eq. 19.  The phonon renormalization is
$$\omega_0^{\prime 2}=\omega_0^2-\sum_{\vec k}{2f_{1\vec k}\alpha_{\vec k+\vec
q_0}(n_{\vec k}-n_{\vec k+\vec q})\over\hbar}.\eqno(D22)$$

\subsection{Effective Electron-electron Interaction}

The electron-electron interaction can be rewritten as
$$H_{ee}=\sum_{\vec k,\vec k^{\prime}}({V_{\vec k}+V_{\vec k^{\prime}}
\over 2})\rho_{\vec k}\rho_{\vec k^{\prime}},\eqno(D23a)$$
with
$$\rho_{\vec k}=W_{\vec k}\psi^{\dagger}_{\vec k+\vec q}\psi_{\vec k},
\eqno(D23b)$$
$$W_{\vec k}={is_x(s_x+c_x)t\over\sqrt{({\Delta\over 2})^2+4t^2(s_x^2+s_y^2)}},
\eqno(D23c)$$
and
$$V_{\vec k}={4\hbar^2g_0^2
\over (E_{\vec k}-E_{\vec k-\vec q_0})^2-\hbar^2\omega_0^2}$$
$$\simeq\cases{-V_0  &if $|E_{\vec k}-E_F|<\hbar\omega_0$,\cr
                          &and $|E_{\vec k-\vec q_0}-E_F|<\hbar\omega_0$,\cr
                    0     & otherwise.\cr}\eqno(D23d)$$
with $V_0=4g_0^2/\omega_0^2$.
Following Balseiro and Falicov\cite{BFal}, this can be decoupled as
$$H_{CDW}=-G_0\sum_{\vec k}\rho_{\vec k}-G_1\sum_{\vec k}^{\prime}\rho_{\vec
k}+{G_0G_1\over\lambda_0},\eqno(D24)$$
with
$$G_0=V_0\sum_{\vec k}^{\prime}<\rho_{\vec k}>,\eqno(D25a)$$
$$G_1=V_0\sum_{\vec k}<\rho_{\vec k}>,\eqno(D25b)$$
where the primed sum means that only states which satisfy the conditions
$|E_{\vec k}-E_F|<\hbar\omega_0$ and $|E_{\vec k-\vec q_0}-E_F|<\hbar\omega_0$
are included.  In the present problem, $E_F=0$.
\par
The resulting Hamiltonian, $H_e+H_{CDW}$ can be
diagonalized by the Bogoliubov-Valatin transformation, yielding quasiparticle
energies
$${\cal E}_{\vec k}^{\pm}={E_{\vec k}+E_{\vec k-\vec q}\over 2}\pm\sqrt{({
E_{\vec k}-E_{\vec k-\vec q}\over 2})^2+G_{\vec k}^2W_{\vec k}^2
},\eqno(D27a)$$
with
$$G_{\vec k}=
       \cases{G_0+G_1     &if $|E_{\vec k}-E_F|<\hbar\omega_0$,\cr
                  G_0     & otherwise.\cr}\eqno(D27b)$$
When $E_{\vec k}=E_{\vec k-\vec q}$, this reduces to
$${\cal E}_{\vec k}^{\pm}\simeq E_{\vec k}\pm |G_{\vec k}W_{\vec k}|,
\eqno(D27c)$$
which should be compared to Eq. C8.

\section{Spin-Orbit Electron-Phonon Coupling}

\par
Earlier suggestions involving charge density waves (CDWs) in the cuprates
were largely abandoned when it was pointed out that the orthorhombic
splitting in the low-temperature orthorhombic (LTO) phase of LSCO leaves the
two in-plane oxygens in symmetrically equivalent positions, and hence should
not split the vHs degeneracy\cite{Poug}.  In fact, there are actually two
independent mechanisms which can lead to a splitting of the vHs degeneracy in
the LTO phase: a dynamic Jahn-Teller effect, where the local symmetry is not
orthorhombic, or spin-orbit coupling, even in a uniform orthorhombic
phase\cite{RMX}.  It would be useful to be able to simplify the
analysis by definitively ruling out one or the other effect.  This has proven
to
be a non-trivial task, as the present estimate makes clear.
\par
Spin-orbit coupling can profoundly change the shape of the Fermi surfaces, by
opening gaps where the bands cross, and thereby splitting the vHs degeneracy in
the dos.  This spin-orbit coupling also leads to a linear coupling to the tilt
modes\cite{Coff,Bones}.  However, the magnitude of these effects is expected to
be small, and hence the effect would be unimportant for driving either the
structural or magnetic transitions.
\par
Here, a simple estimate of $\lambda_{ep}=NV_0$ is presented, which reveals an
underlying subtlety.  The dos can be estimated as
$$N={16\over \pi^2B}ln({B\over\epsilon}),\eqno(E1)$$
where $B$ is the electronic bandwidth and $\epsilon$ is an energy scale, $\sim
T$.  For the three-band model, $B\simeq 8t^2/\Delta$.  Assuming the
renormalized parameters are $t\simeq 0.25eV$, $\Delta\simeq 1eV$, then
$B\simeq 0.5eV$, $\epsilon\simeq 10meV$, and $N\simeq 13eV^{-1}$.
\par
The electron-phonon interaction $V_0$ can be estimated as follows.  The
interaction Hamiltonian can be written\cite{Coff,Bones,RMX}
$$H_{e-ph}\simeq\sum d^{\dagger}p_x\lambda,$$
where the spin-orbit coupling term is $\lambda =\gamma\theta$, with $\theta$
the
octahedral tilt angle,
$$\theta\simeq {2z\over a_0}\equiv \eta (a+a^{\dagger}).$$
Here, $a_0$ is the in-plane lattice constant, $z$ the displacement of the O
out of the CuO$_2$ plane, $a^{\dagger}$ the phonon creation operator, and
$$\eta =\bigl({2\hbar\over m\omega a_0^2}\bigr)^{1/2}\simeq 0.18,$$
with $m$ the mass of an O ion and the bare tilt-mode phonon frequency $\hbar
\omega\sim 5meV$.  The parameter $\gamma$ is, approximately\cite{Bones}
$$\gamma\simeq {\Delta g\over g}t\simeq 25meV$$
with $\Delta g/g\sim 0.1$ the relative g-shift.
Finally,
$$V_0={(\eta\gamma )^2\over\omega}\simeq 4meV,$$
so
$$\lambda_{ep}=V_0N\simeq .05.\eqno(E2)$$
This is too small to drive either the structural or superconducting transition.
However, $\gamma\sim t$, and, up to logarithmic corrections, $N\sim t^2$, so
$\lambda_{ep}\sim t^4$.  Grilli and Castellani\cite{GC} have suggested that,
in the present circumstances, {\it $\lambda_{ep}$ should not be renormalized
by correlation effects}.  This would mean that the {\it bare} value $t=1.3eV$
should be used in Eq. E2, leading to $\lambda_{ep}\simeq 36$, a surprisingly
large value.

\vskip 0.3in
\begin{tabular}{||c||r|r|r||}        \hline
\multicolumn{4}{c}{{\bf Table I: Parameters for Fig. 2}} \\ \hline\hline
\multicolumn{1}{c}{Data Set} &
  \multicolumn{1}{r}{i$\omega_0$} & \multicolumn{1}{r}{$\tilde\omega_0$} &
     \multicolumn{1}{r}{$\tilde\Gamma$} \\
    \hline\hline
(1) & 11.3 & 8.9 & 4060 \\     \hline
(5) & 11.0 & 8.8 & 3880  \\     \hline
(10) & 8.8 & 7.3 & 2550   \\     \hline
(11) & 6.5 & 5.9 & 1500    \\     \hline
\end{tabular}

\bigskip
\bigskip
\centerline{\bf Figure Captions}
\bigskip
{\bf Fig.~1} Phase diagram of LSCO, including the pseudogap transition at $T=T^
*$.  Symbols from Hwang, et al.\cite{Hwa}, short-dashed line = twice the LTO
transition temperature.

{\bf Fig.~2} Doping dependence of LTO transition in
La$_{2-x-y}$Sr$_x$Nd$_y$CuO$
_4$.  Open symbols from Fig. 1 of \cite{Tak}; filled circles from \cite{Buch}.
Lines = theory: solid lines correspond to $y=0$; dashed to $x=0.15$.  Frames a,
b, and c correspond to fit data sets 1, 5, and 10 respectively in Table I (set
11 is indistinguishible from set 10).

{\bf Fig.~3} Phonon modes in LSCO: (a) $\vec q=\vec q_1\equiv(\pi /a,0)$ Cu-Cu
bond stretching mode; (b) $\vec q=\vec q_1$ O-O bond stretching mode; (c) $\vec
q=\vec q_0\equiv(\pi /a,\pi /a)$ Cu-Cu bond stretching mode; (d) $\vec q=\vec
q_0$ O-O bond stretching mode; (e) FE mode; (f) shear strain; (g) possible
mixed
O-O bond stretching, octahedral tilt mode.

{\bf Fig.~4} In-plane Cu-O bond length vs doping, after Radaelli, et
al.\cite{Rad}.  Open circles = 10K data; filled circles = 295K data; $+$'s =
locations of the HTT-LTO transition.
Dot-dashed line = Eq. A12; other lines = guides to the eye.

{\bf Fig.~5} Electronic dispersion associated with static distortions of the
symmetry of the phonon modes in Fig. 3.  The dashed lines show the dispersion
along $\Gamma -Y-M$, whenever this differs from that along $\Gamma -X-M$, while
the dot-(dot-)dashed lines are the dispersion along the line $X (Y)-\bar M
\equiv (\pi /2a,\pi /2a)$.  For ease in comparison, frames e and f are shown
zone-folded into the same $\sqrt{2}\times\sqrt{2}$ supercell as in frames c,d
(beaded lines show original bands).

{\bf Fig.~6} Effect of correlations and polaronic effects on the CuO$_2$
band dispersion in Bi-2212.  Solid circles = data of Dessau, et al.\cite{PE1}.
Fig. 6a: Dotted line = bare parameters (data for LSCO); solid line = parameters
renormalized by correlation effects\cite{RMX}; dashed lines = ghost Fermi
surfaces, due to orthorhombic superlattice.  Fig. 6b: solid and dashed lines =
correlated bands, with additional band narrowing due to polaronic effects.

{\bf Fig.~7} Normalized Fermi velocity, $v_i^*=\hbar\pi v_{Fi}/a$ at a series
of
Fermi levels.  (Renormalized band parameters: $t=0.25eV$, $\Delta =1eV$.)
Fig. 7a: energies below the vHs, $E_{vHs}\simeq 1.207eV$.  Dashed lines, with
increasing $v_x^*$: E = 1.01, 1.02, 1.04, 1.06, 1.08eV; solid lines, with
decreasing $v_x^*$: E = 1.10, 1.12, 1.14, 1.16, 1.18, 1.20eV, and E =
E$_{vHs}$.
Fig. 7b: energies above the vHs.  Solid lines, with increasing $v_x^*$: E =
E$_{vHs}$, 1.21, 1.22, 1.24, 1.26, 1.28eV; dashed lines, with decreasing $v_x
^*$: E = 1.30, 1.32, 1.33eV.
Fig. 7c: $v_x^*$ (solid line) and $E_{\vec k}-E_{\vec k-\vec q_0}$ (dashed
line)
plotted vs. $E_{\vec k}$.

\end{document}